\documentclass[a4paper,11pt]{article}
\pdfoutput=1

\usepackage{jcappub}

\usepackage[T1]{fontenc}
\usepackage{jcappub}
\usepackage{graphicx}
\usepackage{xcolor}

\newcommand{\Trh}{T_{\rm RH}}
\newcommand{\TFO}{T_{\rm FO}}
\newcommand{\Tmax}{T_{\rm max}}
\newcommand{\arh}{a_{\rm RH}}
\newcommand{\aend}{a_{\rm end}}
\newcommand{\am}{a_{\rm m}}
\newcommand{\aFO}{a_{\rm FO}}
\newcommand{\xfo}{x_{\rm FO}}

\usepackage{amsmath,amssymb}
\usepackage{graphicx}
\usepackage{color}

\newcommand{\beq}{\begin{equation}}
\newcommand{\eeq}{\end{equation}}
\newcommand{\bea}{\begin{eqnarray}}
\newcommand{\eea}{\end{eqnarray}}

\title{\boldmath $Z'$ portal dark matter from post-inflationary reheating: WIMPs, FIMPs, and UFOs}

\author[a,1]{Stephen E. Henrich,\note{Corresponding author.}}
\author[b]{Yann Mambrini,}
\author[a]{ and Keith A. Olive}

\affiliation[a]{William I. Fine Theoretical Physics Institute, School of Physics and Astronomy, University of Minnesota \\ Minneapolis, Minnesota 55455, USA}
\affiliation[b]{Universit\'e Paris-Saclay, CNRS/IN2P3, IJCLab, \\ 91405 Orsay, France}

\emailAdd{henri455@umn.edu}
\emailAdd{mambrini@ijclab.in2p3.fr}
\emailAdd{olive@umn.edu}

\abstract{We investigate the production of dark matter (DM) via a heavy $Z'$ mediator during the post-inflationary reheating epoch. In particular, we study 
production from three mechanisms which
are smoothly connected to one another: WIMP-like 
freeze-out, FIMP-like freeze-in, and ultra-relativistic freeze-out (UFO). This is the first systematic study of $Z'$ portal DM which includes UFO. We find that much of the available parameter space for keV to TeV DM lies in the UFO regime for $ 1 \text{ TeV}\lesssim M_{Z'} \lesssim 1 \text{ PeV}$. When the mediator mass $M_{Z'}$ is greater than both the DM mass and the reheating temperature, UFO is a robust mechanism for producing cold DM. Although UFO DM is initially "hot" after freeze-out, it can easily become cold before structure formation if freeze-out occurs during post-inflationary reheating. Compared to standard freeze-in, UFO can accommodate significantly stronger interaction strengths (stronger couplings and/or smaller mediator masses).}

\begin{document}

\begin{flushright}
UMN--TH--4515/25, FTPI--MINN--25/16   \\
December 2025
\end{flushright}

\maketitle
\flushbottom

\section{Introduction}
\label{sec:intro}

Any successful theory of inflation must include an eventual transition from exponential expansion to 
radiation domination. This is most simply accomplished by inflaton decay after inflation ends \cite{dg,afw,grav}. The resulting reheating process is not an instantaneous event. When decays begin, the Universe rapidly heats to a maximum temperature, $T_{\rm max}$, and slowly cools until the radiation bath dominates the total energy density at $T_{\rm RH}$, completing the reheating process. 
The resulting thermal bath must contain the Standard Model (SM) particles,
and all fields 
beyond the SM which interact sufficiently strongly with the SM to achieve equilibrium. Certainly all particles 
with electroweak interactions (and masses below the maximum temperature) 
will be part of the thermal bath initially. 
The thermal out-of-equilibrium scenario, also referred to as thermal freeze out, 
as applied to the relic abundance of a heavy neutral lepton with electroweak interactions is highly predictive \cite{hut}. The cold dark matter (CDM) density is saturated at the value $\Omega h^2 = 0.12$ as inferred from cosmic microwave background (CMB) experiments such as Planck \cite{Planck} for a heavy lepton mass of $m_L \simeq 4 - 5$~GeV, with a slight dependence on whether it is a Dirac or Majorana particle. 
Of course a heavy neutral lepton (or 4th neutrino)
with GeV scale masses must have a mass in excess of $m_Z/2$ to avoid contributing to the width of the $Z$ gauge boson from which we know that there are 2.9963 $\pm$ 0.0074 neutrino states with mass less than $m_Z/2$ \cite{Janot:2019oyi}. However a weakly interacting particle with mass greater than $m_Z/2$ would have a relic abundance far below the needed dark matter density.  
Similarly in the minimal supersymmetric Standard Model (MSSM), there are strong predictions for the properties of the lightest supersymmetric particle (LSP) \cite{gold,ehnos} so long as supersymmetry breaking is at or near the weak scale. Of course to a large extent this predictivity has been lost since the detection of the Higgs boson at 125 GeV, pushing the scale of supersymmetry breaking to the multi-TeV scale, though the thermal out-of-equilibrium mechanism 
still provides for viable dark matter (DM) candidates in the MSSM \cite{Ellis:2022emx,Antusch:2025rbp}.
However if the LSP has only gravitational interactions as in the case of the gravitino, the thermal out-of-equilibrium scenario does not apply as the gravitino was never in thermal equilibrium. Instead, the gravitino is produced by the thermal scattering of SM particles produced during reheating after inflation making its relic abundance proportional to the reheating temperature in standard reheating scenarios \cite{grav,ehnos, kl, grav2}. This mechanism is now typically referred to as freeze-in \cite{fimp,Bernal:2017kxu}.\footnote{Note that in addition to the thermal bath produced by inflaton decay to SM particles,
there is always a relatively high temperature bath produced from the gravitational scattering of the inflaton condensate \cite{Haque:2021mab,cmov}. This bath is important when the reheating process begins and can contribute to the freeze-in production of DM \cite{Henrich:2024rux}.}

Thermal freeze-out occurs whenever the dark matter candidate, $\chi$, comes into 
thermal equilibrium with the SM bath. This requires 
some interaction rate, $\Gamma$, involving the production and annihilation 
of the DM to be fast compared with the 
expansion rate of the Universe characterized by the Hubble parameter, $H$.
Freeze-out occurs when these interactions drop out of equilibrium, 
{\it i.e.} when $\Gamma \lesssim H$ at some temperature $\TFO$. Non-relativistic freeze-out occurs when $\TFO < m_\chi$. Typically for weak scale interactions, we have $\TFO/m_\chi \sim 1/20$ \cite{mybook}. 
In contrast, it is usually assumed that in the freeze-in scenario, the 
DM candidate is never in thermal equilibrium. This may occur if its coupling to 
the SM is extremely weak \cite{fimp}, or if the interactions are mediated by gauge 
(or Higgs) bosons with intermediate or GUT scale masses as in the NETDM scenario \cite{Mambrini:2013iaa,Mambrini:2015vna}. Another scenario referred to as  "freeze-in at stronger coupling" has also been recently explored in \cite{Cosme:2023xpa},
where the authors proposed to replace the feeble coupling 
by the Boltzmann suppression of the particles in the bath, $e^{-\frac{m_\chi}{T}}$, 
if the mass of the DM lies above the reheating temperature.

However, the production of DM is not limited to these two mechanisms, non-relativistic freeze-out (FO) and freeze-in (FI). 
Indeed, depending on the details of reheating (to be discussed in more detail in Section \ref{reheating}), the DM candidate may come into equilibrium and freeze-out before reheating is complete or the DM candidate may freeze-out (permanently) while it is still ultra-relativistic \cite{Henrich:2025gsd,HMO2}. 
This mechanism is called ultra-relativistic freeze out, or UFO.
This is for instance the case for SM neutrinos which undergo UFO when their electroweak 
interactions decouple at a temperature $\TFO \sim 1 $~MeV. These neutrinos would saturate the density of matter if their sum totaled about 11 eV, but remain hot DM 
candidates and are therefore excluded because of their negative impact on 
structure formation. However, as we will see, 
the UFO mechanism becomes viable if freeze-out occurs 
{\it during reheating}
and the DM candidates have a chance to cool down so that they act as CDM by the time structures in the Universe begin to form. 

While it is often convenient to characterize DM production as either freeze-out (FO or UFO) or freeze-in, there is in reality a continuum of possibilities 
which lead to the correct abundance. The relic density clearly depends on the DM mass and the interaction strength  between the DM and the SM. This in turn may depend on the 
size of a fundamental coupling, $g$ (perhaps a gauge coupling) or the mass of a 
mediator, $M$ (though often there is a degeneracy since the rate may depend on $g/M$). As we will discuss, the abundance may also depend on the reheating 
temperature and the details 
of the reheating process. Indeed, the freeze-out and/or freeze-in mechanisms 
may be important for either 
fermionic or scalar DM with either scalar or vector mediators.  In this work, we propose a unified picture of the DM production process assuming fermionic DM with 
interactions mediated by vector exchange. Our results are thus mainly sensitive to 
three parameters, $M$, $m_\chi$, and $\Trh$. As these are varied, we may 
continuously move from FI to UFO to FO as the DM mass increases.\footnote{There 
are of course several other mechanisms for the production of DM which do not involve the presence of a thermal bath, like the direct decay of the inflaton \cite{KMO}, gravitational inflaton scattering \cite{Henrich:2024rux,cmov}, or spectator fields produced during inflation \cite{Garcia:2025rut}. 
We do not consider them here.}

In particular, we investigate the production of dark matter (DM) via $Z'$-portal 
interactions during post-inflationary reheating.
$Z'$-portal DM models arise naturally in well-motivated, 
simple extensions of the Standard Model involving an extra $U(1)'$ gauge 
symmetry \cite{Langacker:2008yv}. While $Z'$ portal DM is popular in minimal BSM 
models, heavy vector portals also arise in GUT models such as SO(10), which naturally contain a $U(1)_{B-L}$ gauge symmetry \cite{Mambrini:2016dca,Mambrini:2015vna}.

Nearly all previous studies investigating $Z'$-portal DM exclusively consider DM production in the radiation dominated era. Here, we consider the possibility that $Z'$ portal DM may be produced {\it during} the 
reheating era preceding radiation domination. For a given production rate, the resulting abundance of DM may be very different.
Indeed, there are several new phenomena which emerge when non-instantaneous reheating is taken into account: 
\begin{itemize}
\item There is the possibility that the DM mass $m_{\chi}$ may be greater than the reheating temperature, $T_{\rm RH}$. Typically, a thermal bath with temperature $T_{\rm max}$ is established quickly after exponential expansion ends \cite{Scherrer:1984fd,Giudice:2000ex,GMOP,KMO,GKMO1,GKMO2}. Therefore particle production with $T > m_\chi > \Trh$ can be important \cite{GKMO1,Henrich:2024rux}.   This alters the observed relic abundance compared to the standard production during radiation domination, because the expansion rate of the universe during reheating is generally quite different than the expansion during radiation domination. This means that there can be significant dilution of the co-moving DM number density $n_{\chi}$ after freeze-out or freeze-in production. This non-trivially affects the final DM abundance. 

\item There is also the possibility for the DM abundance 
to be sensitive to either a characteristic UV scale or IR scale \cite{GMOP,KMO,GKMO1}. While this possibility 
also exists during radiation domination, the two scales are different when production occurs during reheating,
where the UV scale is typically the maximum temperature reached by the thermal bath, $T_{\rm max}$. 
The IR scale can be one of two possibilities: either 1) the reheating temperature (when $m_{\chi} < T_{\rm RH}$) 
or 2) the DM mass $m_{\chi}$ (when $m_{\chi}> T_{\rm RH}$). This distinct scale-dependence introduces a rich 
new set of possibilities for DM phenomenology. 

\item Depending on the specific reheating model, there are multiple possible ways that the Hubble parameter and consequently the SM radiation bath can evolve during the reheating period, leading to distinct outcomes for the final abundance. This will be discussed in greater detail in Section \ref{reheating}.

\item Finally, in addition to the possibility of $m_{\chi}>T_{\rm RH}$, the distinct scale sensitivity, and distinct 
evolutions of the SM bath, we have recently shown that 
reheating also enables the possibility that DM can undergo 
(ultra)relativistic freeze-out (UFO) and still have sufficient time to cool to become cold DM before structure formation
\cite{Henrich:2025gsd}. 
We discuss this surprising new possibility in more detail below. 
\end{itemize}
All of these distinctive features of DM production during reheating can non-trivially impact approaches for DM detection and the associated exclusion plots.

To the best of our knowledge, only one other study has examined $Z'$ portal DM during reheating \cite{Belanger:2024}. In this work, the authors considered $Z'$ portal DM production in the 
setting of low reheating temperatures and a low inflationary scale. Considering 
SM fermions with a vector 
coupling to $Z'$ with strength $V_f$, the authors found that 
current detection experiments are already probing parameter 
space associated with production during the reheating era.  Two benchmark scenarios were considered, namely $M_{Z'}=1$~GeV with $V_{f}=10^{-5}$, and $M_{Z'}=500$~GeV with $V_{f}=10^{-4}$. 
They also considered a low Hubble parameter at the end of inflation,
$H_{\rm end}=0.1$~GeV. This type of model allows for larger than usual couplings for the freeze-in mechanism, and therefore a potentially observable FIMP~\footnote{We note that in a radiation-dominated Universe, the effective coupling $\lambda_{eff}$ for a FIMP which allows a present relic abundance of 
$\frac{n_\chi(T_0)}{n_\gamma(T_0)} \simeq 10^{-9}$ is of the order of $\lambda_{eff}\sim 10^{-11}$, rendering $\chi$ undetectable, even for future generations of detectors.}, along with smaller than usual couplings for WIMPs. We note that other studies have also investigated both WIMPs and FIMPs during non-instantaneous reheating \cite{Haque:2023yra,Silva-Malpartida:2023yks}. 

In addition to this work, an interesting alternative
to the standard FIMP scenario was recently proposed \cite{Arcadi:2024}. This unconventional production mechanism is based on the idea of
freeze-in at stronger coupling.
In the original proposition \cite{Cosme:2023xpa}, it was 
assumed that the Standard Model radiation bath never reached temperatures greater than the DM mass, leading to 
Boltzmann-suppressed DM production. In effect, the weak coupling typically associated with a FIMP is substituted by a Boltzmann factor $e^{-\frac{m_\chi}{T}}$, which necessitates
a relatively low reheating temperature if we want to exploit candidates in a mass range  observable by current direct detection experiments, $m_\chi \lesssim 100$ GeV. 
As a consequence, larger couplings between DM and SM are allowed,
while remaining within the freeze-in paradigm, rendering the models observable in direct detection experiments.

In both scenarios described above, a sufficiently low reheating temperature is necessary to ensure the correct relic density. 
It should be kept in mind that a low reheating temperature (e.g. below the electroweak scale) can be problematic for the generation of a baryon asymmetry through leptogenesis, for example.  Furthermore, these two studies, as in most FI scenarios, neglect the initial conditions set up at the end of inflation. In other words, any coupling between the particle responsible for reheating (inflaton, reheaton) and dark matter is ignored. This assumes a certain form of exact symmetry that would need to be justified a priori. In the absence of a symmetry such a coupling must exist, at least at the loop level \cite{KMO}. 
A third approach has been recently proposed that does not require a low reheating temperature.
This mechanism is known as ultra-relativistic freeze-out (UFO), which is a newly 
recognized mechanism for generating cold DM when UFO occurs during the reheating era \cite{Henrich:2025gsd,HMO2}.

The possibility of UFO DM is not new. In fact, 
it may be the first mechanism proposed 
to obtain a viable DM candidate in the 60's, considering the case of a light massive neutrino \cite{Gershtein:1966gg,Cowsik:1972gh,Szalay:1974jta,SS}.
The idea is similar to that of SM neutrino decoupling from the thermal bath. But instead of considering a radiation-dominated 
Universe, we assume that decoupling takes place during reheating. Even though the process is physically the same, the 
fact that there is a continuous injection of entropy into the thermal bath during reheating completely changes the dynamics.
Unlike the freeze--in scenario studied in \cite{Belanger:2024}, in the UFO case, 
DM {\it is in thermal equilibrium before its decoupling}, 
effectively eliminating any dependence on initial conditions.
It leaves thermal equilibrium because its interactions do not allow it to remain in equilibrium with the particles of the Standard Model. However, unlike the classic case of SM neutrinos, 
the continuous injection of entropy between $\TFO$ and $\Trh$ allows for the effective dilution of dark matter to obtain the correct present-day relic abundance. Furthermore, unlike WIMP-like FO, significant out-of-equilibrium production (similar to FI) can occur after UFO \cite{Henrich:2025gsd,HMO2}. Importantly, the UFO mechanism does not rely upon \textit{ad hoc} entropy injection, but rather it simply uses the entropy injection which is already required in standard inflationary cosmology to reheat the universe. Although the UFO mechanism has been well explained and explored in \cite{Henrich:2025gsd,HMO2} in a general context, it has not yet been applied to a specific microscopic model. Furthermore, as we will show, UFO continuously interpolates between the FI and FO mechanisms.

  The UFO mechanism has previously been neglected in $Z'$ portal studies for several reasons. First, because prior studies typically investigate DM production during radiation domination, the UFO mechanism would lead to hot relic DM and therefore be not viable. Second, prior studies \cite{Belanger:2024} which did consider $Z'$ portal DM during the reheating era did not consider $Z'$ masses large enough for the UFO regime to become visible. 
This may seem counter-intuitive, but once again, it is useful to recall the case of the SM neutrino. It is only at 
temperatures {\it below} the mass of $Z$ that the neutrino can freeze-out. In other words, it is only because $Z$ is sufficiently 
massive that the neutrino was able to decouple from the thermal 
bath. This is directly related to the dependence of the $\nu$-thermal bath interaction rate, $\Gamma_\nu \sim \frac{T^5}{M_Z^4}$, compared to the Hubble rate $H(T)\sim \frac{T^2}{M_P}$.           The same applies to UFO-type dark matter, whose 
intermediary is a massive $Z'$ vector.

To be more precise, we have previously shown \cite{Henrich:2025gsd,HMO2} that the
UFO scenario appears
if the thermally averaged cross section has a sufficiently steep temperature dependence
\beq
\langle \sigma v \rangle \propto \frac{T^n}{\Lambda^{n+2}} \, ,
\eeq 
where the lower limit on the exponent $n$ depends on the details of reheating.
For example, UFO requires that in the relativistic regime, the production and annihilation
rates, $\Gamma_{\rm rel}=\langle \sigma v \rangle n_{{\rm eq},\chi}$, where $\langle \sigma v \rangle$ is the thermally averaged cross section, and $n_{{\rm eq},\chi}$ is the equilibrium number density of $\chi$, 
have a {\it steeper} temperature dependence than the Hubble rate. 
This is to ensure relativistic freeze-out, or 
$\TFO>m_\chi$.
So if $\Gamma(T) \propto T^{\gamma_{1}}$ and $H(T) \propto T^{\gamma_{2}}$, then UFO requires $\gamma_{1}>\gamma_{2}$. 
For standard reheating, where the inflaton undergoes matter-like oscillations, the temperature dependence of the Hubble 
parameters is $H(T) \propto T^4$
\cite{GKMO1,GKMO2,Garcia:2021iag}, so $\gamma_2=4$ during 
reheating and we expect $\gamma_1 = n+3$ when $m_{\chi}\ll T$ 
such that the DM is relativistic and the number density simply goes as $n_{{\rm eq},\chi} \propto T^3$. The thermally averaged cross section for $Z'$ portal 
DM production via $2\rightarrow 2$ annihilations ($\bar{f}f\rightarrow \bar{\chi}\chi$) where $M_{Z'}>T$ is characterized 
by $n=2$ temperature dependence, which implies $\gamma_{1}=5$. 
Thus, for \textit{heavy} $Z'$ portal DM, UFO is possible. However, for a light $Z'$, corresponding to a long-range interaction, where $T\gtrsim M_{Z'}$, the 
temperature dependence of the thermally averaged cross section 
is instead characterized by $n=-2$, such that $\gamma_{1}=1$, 
UFO is not possible since $\gamma_{1}<\gamma_{2}$. This 
explains why previous studies of $Z'$ portal DM during reheating, as in \cite{Bhattacharyya:2018evo}, 
did not consider the UFO regime.

Our study is distinct from prior works in several important ways. First, we directly tie the initial 
conditions at the beginning of the reheating period to well-motivated models of inflation which are constrained to be consistent with CMB observables. Though our results do not depend on the specific inflation model, we do have in mind something close to the Starobinsky model of inflation \cite{Staro}. In these cases the scale of inflation is fixed by the amplitude of density perturbations as determined by Planck \cite{Planck} and leads to $H_{\rm end }= \frac{\sqrt{\rho_{\rm end}}}{\sqrt{3} M_P} \approx 10^{12}-10^{13}$~GeV, where $\rho_{\rm end}$ is the inflaton energy density at the end of inflation and $M_P = 2.4 \times 10^{18}$~GeV is the reduced Planck mass. 
Second, we study a much broader range of heavy $Z'$ masses ($10^{4}\leq M_{Z'}\leq 10^{10}$~GeV). 
Third, we include axial vector as well as vector couplings. Fourth, we extend our study to the entire viable range of 
reheating temperatures, {\it i.e.} $4$~MeV $\lesssim T_{\rm RH}\lesssim (H_{\rm end} M_P)^\frac12$.
We do note that models proposing freeze-out in the relativistic regime have been discussed in the context of a self-interacting hidden sector \cite{Hambye:2019tjt,Hambye:2020lvy,Coy:2021ann,Coy:2024itg,ArcadiLebedev:2019,SterileNeutrinoFOLebedev}. While these dark matter candidates 
do not come into equilibrium with the SM, their self-interactions establish a temperature distinct from the SM bath and have the possibility of freezing out from a secluded sector while relativistic. For relativistic treatments of freeze-in, see \cite{FreezeInUltra1,FreezeInRel1,FreezeInRel2}.

Lastly, and perhaps most distinctively, we study not only WIMP-like 
freeze-out and FIMP-like freeze-in, but also UFO.
Therefore, our study is significantly more general in some respects than prior works, and crucially includes the heretofore unexamined UFO mechanism for $Z'$-portal DM. Note also that freeze-in with
a $Z'$-portal DM has been thoroughly studied during the radiation-dominated era \cite{Cosme:2021}. In this work, both 
vector and axial vector couplings were examined, and both $Z'$ decays and $Z'Z' \rightarrow \bar{\chi}\chi$ processes were 
taken into account. The authors studied $Z'$ mediators with masses ranging from MeV to PeV scales, and found the current 
experiments are already probing parameter space associated with FIMPs. Collider constraints on $Z'$ portal DM have been studied in the context of the $U(1)_{B-L}$ model \cite{Klasen2016, Okada2018}, and were also considered for the freeze-in at stronger coupling scenario \cite{Arcadi:2024}. But none
of these works included the UFO contribution.

In what follows, we will first briefly review the
standard reheating process after inflation in Section \ref{reheating}. This is followed by our calculation of relevant DM production rates in Section \ref{rates}. Our analytic (approximate) solutions of the Boltzmann equations for the DM relic density is provided in Section \ref{relab}.
Our main results are presented in Section \ref{results} as contours in the $(\Trh,m_\chi)$ plane consistent with the observed relic abundance for fixed $M_{Z'}$, for both vector and axial vector interactions. Our conclusions are given in Section \ref{summ}.

\section{Approach}

\subsection{Post-inflationary reheating}
\label{reheating}

Much of our analysis is focused on the production of dark matter during the reheating epoch. 
Nevertheless, our results do not depend on the details of any particular model of inflation beyond considering an inflationary scale of order $10^{13}$~GeV. This can be easily motivated by the Starobinsky model \cite{Staro} (or any related model) in which the potential takes the form
\beq
V \; = \; \frac34 M^2 M_P^2 \left(1 - e^{-\sqrt{\frac23} \frac{\phi}{M_P}} \right)^2 \, ,
\label{starpot}
\eeq
where the Hubble parameter during inflation is set by the inflaton mass $H_{\rm I} = m_\phi/2 = M/2$. 
This scale is fixed by the amplitude of scalar fluctuations
\beq
A_s \; = \; \frac{3 M^2}{8\pi^2 M_P^2} \sinh^4 \left(\frac{\phi_*}{\sqrt{6}M_P} \right) \, , 
\label{AsStaro}
\eeq
in the Starobinsky model. Planck \cite{Planck} has determined $A_s = 2.1 \times 10^{-9}$, 
and $\phi_* \simeq 5.35 M_P$ is the value of the inflaton field prior to the final $N_*\sim 55$ $e$-folds of inflation.\footnote{The precise value of $\phi_*$ (and $N_*$) are 
mildly dependent on the reheating temperature \cite{EGNO5,egnov}.} This fixes $m_\phi = 1.25 \times 10^{-5} M_P \simeq 3 \times 10^{13}$~GeV.  Inflation ends when $\phi = \phi_{\rm end} \simeq 0.6 M_P$, determined when $\ddot a = 0$, where $a$ is the cosmological scale factor. 
Subsequently, 
the inflaton begins a series of harmonic oscillations with 
frequency, $m_\phi$. 

As oscillations begin, inflaton decay products start to populate the
thermal bath.
The energy density stored in the inflaton condensate is determined from the inflaton equation of motion so that 
\beq
{\dot \rho}_\phi + (3 H +\Gamma_\phi) \rho_\phi = 0 \, ,
\eeq
where the inflaton decay rate can be parameterized by $\Gamma_\phi = y^2 m_\phi/8\pi$, where $y$ is an effective Yukawa-like coupling between the inflaton and SM fields. For $\Gamma_\phi \ll H$, 
$\rho_\phi \propto a^{-3}$ and the inflaton condensate redshifts as non-relativistic matter. The radiation density is determined from 
\beq
{\dot \rho}_{\rm R} + 4 H \rho_{\rm R} = \Gamma_\phi \rho_\phi
\label{Eq:rhor}
\eeq
and the radiation density scales as $\rho_{\rm R} \sim T^4 \sim a^{-3/2}$. As soon as oscillations begin, the thermal bath is established and reaches a maximum temperature. 
Solving Eq.~(\ref{Eq:rhor}) exactly, we can compute the maximum temperature $\Tmax$ reached by the bath \cite{GKMO2}
\beq
\alpha T_{\rm max}^4 = \frac{\sqrt{3}}{32\pi} \left(\frac38 \right)^\frac35 y^2 m_\phi M_P \rho_{\rm end}^\frac12 \, ,
\eeq 
where $\alpha = g^* \pi^2/30$ for $g^*$ relativistic degrees of freedom\footnote{We assume the SM value $g^* = 427/4$ which is valid for temperatures above the top quark mass. For lower reheating temperatures, this factor must be adjusted but we do not explicitly include factors of $g$ for lower temperatures in our analytical expressions.}, and  $\rho_{\rm end} = \frac32 V(\phi_{\rm end}) \simeq (7 \times 10^{15}~{\rm GeV})^4$.
The solution of Eq.~(\ref{Eq:rhor}) also determines the evolution of the thermal bath, 
\beq
\rho_{R}=\alpha T^4 = \alpha T_{\rm RH}^4 \left(\frac{a_{\rm RH}}{a}\right)^{3/2} \, ,
\label{Eq:rhoRad}
\eeq
 and $a_{\rm RH}$ is the scale factor at reheating  which is determined from the condition $\rho_\phi(\arh)=\rho_R(\arh)=\alpha \Trh^4$, and gives
\cite{GKMO2}
\beq
\alpha \Trh^4 = \frac{3}{400\pi^2}y^4 m_\phi^2 M_P^2 \, , 
\eeq
where we used the fact that, in the Starobinsky model, the inflaton field behaves as non-relativistic matter during the oscillatory phase.

The Hubble parameter during the reheating era is given by 
\beq
H(a) = \frac{1}{\sqrt{3}M_{P}}\sqrt{\rho_{\phi}(a)+\rho_{\rm R}(a)}\approx\frac{\sqrt{\rho_{\phi}(a)}}{\sqrt{3}M_{P}}=\frac{\sqrt{\rho_{\rm end}}}{\sqrt{3}M_{P}} \left( \frac{a_{\rm end}}{a}\right)^{3/2},
\label{Eq:Hubble}
\eeq
where $\rho_{\rm end}$ is the energy density stored in the inflaton at the end of inflation and $a_{\rm end}$ is the scale factor at that time.

To get an idea of the expected order of magnitude, for $y = 10^{-5}$, we obtain $T_{\rm max} \simeq 3 \times 10^{12}$~GeV and $\Trh \simeq 6 \times 10^9$~GeV.  UV production of DM will occur at 
or near $T_{\rm max}$ and IR production at or near  either $\Trh$ or $m_\chi$. Note also that for $y \lesssim 3 \times 10^{-6}$, the gravitational process dominates the early phase of reheating. 
The gravitational bath due to inflaton scattering exchanging a graviton $h_{\mu \nu}$ 
sets the maximum temperature, $T_{\rm max}^h$ to \cite{cmov}
\beq
\alpha \left({T_{\rm max}^h}\right)^4 =\frac{\sqrt{3}}{144 \pi} \left(\frac89 \right)^8 \frac{m_\phi \rho_{\rm end}^\frac32}{M_P^3} \simeq 2\times 10^{12}~{\rm GeV}\,,
\eeq
and is of course independent of $y$. For the role of the gravitational bath in freeze-in processes, see \cite{Henrich:2024rux}.

\subsection{Dark matter production rates}
\label{rates}

We consider a general $Z'$ portal interaction where the relevant portion of the Lagrangian is
\begin{equation}
\mathcal{L} \supset \bar{\chi}\gamma^{\mu}(V_{\chi}+A_{\chi}\gamma_{5})\chi Z'_{\mu} +\sum_{f}\bar{f}\gamma^{\mu}(V_{f}+A_{f}\gamma_{5})fZ'_{\mu} - \frac{1}{2}M_{Z'}^2Z'^{\mu}Z'_{\mu}-m_{\chi}\bar{\chi}\chi,
\end{equation}
where $V_{i}$ and $A_{i}$ are the vector and axial vector couplings for particle
\textit{i}, $\chi$ is a Dirac fermion, and the sum runs over all Standard Model (SM) fermions $f$ that are charged under the additional $U(1)'$ gauge group.

The DM production rate, $R_{\chi}(T)$, for the process $\bar{f}f\rightarrow\bar{\chi}\chi$ can be expressed as a thermal average over the initial state energies, 
\begin{equation}
R^{\bar{f}f}_{\chi}(T)=\sum_{f}\frac{1}{1024\pi^6}\int f(E_{1})f(E_{2}) \sqrt{1-\frac{4m_{\chi}^2}{s}}E_{1}~\text{d}E_{1}E_{2}~\text{d}E_{2}~\text{d}\cos\theta_{12}\int |\mathcal{M}_{\bar{f}f\rightarrow\bar{\chi}\chi}|^{2}~\text{d}\Omega_{13},
\label{Eq:rate}
\end{equation}
where the sum runs over all SM fermions that are charged under the additional $U(1)^{'}$, and $f(E_{i})$ is the Fermi distribution function for initial state SM particle $i$ \cite{cmov,Clery:2022wib}. The infinitesimal solid angle $\text{d} \Omega_{13}$ is defined to be 
\begin{equation}
\text{d}\Omega_{13}=2\pi~\text{d}\cos\theta_{13}
\end{equation}
where $\theta_{12}$ and $\theta_{13}$ are the angles between the momenta $\vec{p}_{1,2}$ and $\vec{p}_{1,3}$ respectively.

In Eq.~(\ref{Eq:rate}), $|\mathcal{M}_{\bar{f}f\rightarrow\bar{\chi}\chi}|^{2}$
is the spin-averaged squared matrix element for the process $\bar{f}f\rightarrow\bar{\chi}\chi$. We find the following contributions:
\begin{align}
|\overline{\mathcal{M}_{\rm vv}}|^{2} &= \frac{2 V_{f}^2 V_{\chi}^2}{(s-M_{Z'}^2)^2+\Gamma_{Z'}^2M_{Z'}^2}[s^2+2st+2(m_{\chi}^2+m_{f}^2-t)^2] \, , \label{Eq:Mvv} \\
|\overline{\mathcal{M}_{\rm va}}|^{2} &= \frac{2}{(s\!-\!M_{Z'}^2)^2\!+\!\Gamma_{Z'}^2M_{Z'}^2}\left[(V_{f}^2A_{\chi}^2\!+\!V_{\chi}^2A_{f}^2) (s^2+2st+ 2(m_{\chi}^2\!+\!m_{f}^2\!-\!t)^2) \nonumber \right. \\ & \left. -4V_{f}^2A_{\chi}^2m_{\chi}^2 (s+2m_{f}^2)-4V_{\chi}^2 A_f^2m_{f}^2 (s+2m_{\chi}^2) \nonumber \right. \\ & \left. -V_{f}V_{\chi}A_{f}A_{\chi}(4s^{2}+8st-8s(m_{\chi}^2+m_{f}^2))\right] \, , \label{Eq:Mva} \\
|\overline{\mathcal{M}_{\rm aa}}|^{2} &= \frac{2 A_{f}^2A_{\chi}^2}{(s-M_{Z'}^2)^2+\Gamma_{Z'}^2M_{Z'}^2} \Bigr[s^2+2st+2(m_{\chi}^2+m_{f}^2-t)^2-4s(m_{\chi}^2 +m_{f}^2)  \nonumber \\ & +16m_{f}^2m_{\chi}^2-16\frac{sm_{f}^2m_{\chi}^2}{M_{Z'}^2}+8\frac{s^{2}m_{f}^2m_{\chi}^2}{M_{Z'}^4} \Bigr] \label{Eq:Maa} \, .
\end{align}

To compute the rate, it is useful to relate the Mandelstam variables, $s$ and $t$ to the angles $\theta_{12}$ and $\theta_{13}$ and the DM mass, $m_\chi$ assuming that we can neglect the SM fermion masses.
\begin{align}
s&=2E_{1}E_{2}(1-\cos\theta_{12}) \,,\nonumber
\\
t&=\frac{s}{2}\left(\sqrt{1-\frac{4m_{\chi}^2}{s}}\cos\theta_{13}-1\right)+m_{\chi}^2\,. \nonumber \\
u&=2m_{\chi}^2-s-t
\label{Eq:st}
\end{align}
In the regime where $m_{f},m_{\chi} \ll T \ll M_{Z'}$, inserting the expressions of $s$
and $t$ from Eq.~(\ref{Eq:st}) into the expression for
$|\mathcal{M}_{\bar{f}f\rightarrow\bar{\chi}\chi}|^{2}$, we find for the thermally averaged rate (\ref{Eq:rate})
\begin{equation}
R^{\bar{f}f}_{\chi}=  \frac{49\pi^3 \Sigma_{\rm tot}}{259200}\frac{T^8}{M_{Z'}^4}\,,
\label{Eq:rateLimit1}
\end{equation}
where we have defined 
\beq
\Sigma_{\rm tot} = \left( \sum_{f}g_{\rm eff,f}^4 \right)
\label{Eq:sigmatot}
\eeq
and 
\beq
g_{\rm eff,f}=(V_{f}^2V_{\chi}^2\!+\!V_{f}^2A_{\chi}^2\!+\!V_{\chi}^2A_{f}^2\!+\!A_{\chi}^2A_{f}^2)^{\frac{1}{4}}\,,
\label{Eq:geff}
\eeq
and the sum is over all Standard Model fermions $f$ charged under $U(1)'$. 

In the same limit where the SM fermion masses are neglected, but $m_{\chi} \lesssim T$ is kept, 
we obtain: 
\begin{align}
R^{\bar{f}f}_{\chi,vv}&=\left( \sum_{f}V_{f}^2 V_{\chi}^2 \right)\frac{T^8}{M_{Z'}^4}\left[ \frac{49\pi^3}{259200}+\frac{3\zeta(3)^2}{16\pi^5}\left(\frac{m_{\chi}}{T}\right)^2-\frac{1}{4608\pi}\left(\frac{m_{\chi}}{T}\right)^4\right] 
\label{Eq:rateVV}\,,
\\
R^{\bar{f}f}_{\chi,va}&=\left( \sum_{f}V_{f}^2 A_{\chi}^2\!+\!A_{f}^2 V_{\chi}^2 \right)\frac{T^8}{M_{Z'}^4}\left[\frac{49\pi^3}{259200}+\frac{3\zeta(3)^2}{16\pi^5}\left(\frac{m_{\chi}}{T}\right)^2-\frac{1}{4608\pi} \left(\frac{m_{\chi}}{T}\right)^4\right] \nonumber \\ &-\left( \sum_{f}V_{f}^2 A_{\chi}^2 \right)\frac{9\zeta(3)^2}{32 \pi^5}\frac{T^6 m_{\chi}^2}{M_{Z'}^4}\,,
\label{Eq:rateVA}
\\
R^{\bar{f}f}_{\chi,aa}&=\left( \sum_{f}A_{f}^2 A_{\chi}^2 \right)\frac{T^8}{M_{Z'}^4}\left[ \frac{49\pi^3}{259200}-\frac{3\zeta(3)^2}{32\pi^5}\left(\frac{m_{\chi}}{T}\right)^2-\frac{1}{4608\pi}\left(\frac{m_{\chi}}{T}\right)^4\right]\,.
\label{Eq:rateAA}
\end{align}
The total rate can be written by combining the above contributions. 
We find the total rate to be
\begin{align}
R^{\bar{f}f}_{\chi,\rm tot}&=\left[\frac{49\pi^3\Sigma_{\rm tot} }{259200}\right]\frac{T^8}{M_{Z'}^4}+\left[\sum_f (V_f^2 V_\chi^2\!+\!A_f^2 V_\chi^2)\frac{3\zeta(3)^2}{16\pi^5}\!-\!\sum_f (V_f^2 A_\chi^2\!+\!A_f^2 A_\chi^2)\frac{3\zeta(3)^2}{32 \pi^5}\right]\frac{T^6 m_{\chi}^2}{M_{Z'}^4} \nonumber \\ &-\left[\frac{\Sigma_{\rm tot}}{4608\pi}\right]\frac{T^4 m_{\chi}^4}{M_{Z'}^4}\,.
\end{align}
Alternatively, we can consider the limit where $T \gg M_{Z'},m_{\chi},m_{f}$, we then find
\begin{align}
R^{\bar{f}f}_{\chi,\rm tot}&=\left[\frac{\Sigma_{\rm tot}}{13824\pi}\right]T^4 \, .
\label{Eq:rateHighTemp}
\end{align}
Lastly, we can consider the rate in the limit where $T \lesssim m_\chi$. In this limit, the Boltzmann suppression factor must 
be taken into account, since only the fraction of 
particles with energy above $m_\chi$ will be able to produce DM. In this case the rate can be expressed as
\beq
R^{\bar{f}f}_{\chi,\rm tot} = \frac{ 16 \pi \left( \sum_{f}V_{f}^2 V_{\chi}^2 \right) m_\chi^2 }{M_{Z'}^4}  n_{\rm eq}^2 \, ,
\eeq
where 
\beq
n_{\rm eq}= \frac{g_\chi}{2\pi^2} \int p^2 dp f(E) =\begin{cases} \frac{3 \zeta(3)  g_\chi}{4\pi^2} T^3 \qquad T \gg m_\chi \,,
\\ \frac{g_{\chi}}{(2\pi)^{3/2}}(m_{\chi}T)^{3/2}e^{-m_{\chi}/T} \qquad  T\ll m_\chi\,,
\label{Eq:neq}
\end{cases}
\eeq
with $E=\sqrt{p^2+m_\chi^2}$, and $g_\chi$ internal degrees of freedom of $\chi$. \footnote{Note that while in this section we have derived multiple analytic expressions for the rate $R^{\bar{f}f}_{\chi}$ in various limits, in our results section below we have computed $R^{\bar{f}f}_{\chi}$ numerically rather than resorting to these analytic expressions.}

In addition to the $\bar{f}f\rightarrow\bar{\chi}\chi$ annihilation process, we must also include the $Z'\rightarrow\bar{\chi}\chi$, $Z'\rightarrow\bar{f}f$, $Z'Z'\rightarrow\bar{\chi}\chi$, and $Z'Z'\rightarrow\bar{f}f$ processes.
For the $Z'Z'\rightarrow \bar{\chi}\chi$ process, we take the expression for the matrix element squared given in \cite{Cosme:2021,Gray:2020thesis}.
The rates, $R^{Z'Z'}_{\chi}$ for the $Z'Z' \to \bar{\chi}\chi$ processes can be obtained from Eq.~(\ref{Eq:rate})
with the substitution of that matrix element instead of Eqs.~(\ref{Eq:Mvv} - \ref{Eq:Maa}) and compute the result numerically to obtain the corresponding rate as a function of temperature.

 For the decay 
processes, $Z'\rightarrow \bar{\chi}\chi/\bar{f}f$, we find
\begin{equation}
|\overline{\mathcal{M}}|_{Z'\rightarrow\bar{\chi}\chi/\bar{f}f}^{2}=\frac{4}{3}\left[V_{\chi/f}^2(M_{Z'}^2+2m_{\chi/f}^2)+A_{\chi/f}^2(M_{Z'}^2-4m_{\chi/f}^2)\right] \, .
\label{decayAmp}
\end{equation}
In this case,
the thermally averaged decay rate can readily be found as follows. First we determine the rest frame decay 
rate, $\Gamma_0$ using the generic expression
\begin{equation}
\Gamma_{0}=\frac{1}{2M_{Z'}}\int\frac{d^3p_1}{(2\pi)^3 2E_{1}}\frac{d^3p_2}{(2\pi)^3 2E_{2}}(2\pi)^4 \delta^{4}(P-p_1-p_{2})\overline{|\mathcal{M}_{{Z'}\rightarrow \bar{\chi}\chi}|}^2\,.
\end{equation}
Using the amplitude squared we found above the decay process in Eq.~(\ref{decayAmp}), we find
\begin{equation}
\Gamma_{0}=\frac{M_{Z'}}{12\pi }\sqrt{1-\frac{4m_{\chi/f}^2}{M_{Z'}^2}}\left[V_{\chi/f}^2\left(1+\frac{2m_{\chi/f}^2}{M_{Z'}^2}\right)+A_{\chi/f}^2\left(1-\frac{4m_{\chi/f}^2}{M_{Z'}^2}\right)\right]\,.
\end{equation}
The corresponding thermally averaged decay rate can then be found as follows:

\begin{equation}
\langle \Gamma \rangle = \frac{\int d^3 p f(p)\Gamma_{0}\frac{M_{Z'}}{E}}{\int d^3 p f(p)}\,,
\end{equation}

\noindent where in this case $f(p)$ is the Bose distribution function and we have used the fact that the decay rate in the plasma rest frame is subjected to a momentum-dependent time dilation $\Gamma(E)=\Gamma_{0}\frac{M_{Z'}}{E}$. The thermally averaged decay rate per unit volume is then 
\beq
R^{Z'}_\chi = \langle \Gamma \rangle n^{\rm eq}_{Z'} \, .
\eeq
The decay rate is smaller because the lifetime of the particle in the thermal frame is larger than the rest frame lifetime. We additionally note that while many studies of $Z'$ portal DM use the narrow width approximation in the resonance region, we do not use this approximation here since we consider coupling of $Z'$ to all SM fermions with $V_f=1$, which corresponds to a fairly broad decay width.

\subsection{Relic abundance}
\label{relab}

Next, we will determine the relic abundance of DM via freeze-in, (ultra)relativistic freeze-out, or WIMP-like freeze-out during non-instantaneous reheating. Our starting point is the Boltzmann equation describing the evolution of the DM number density \cite{Cosme:2021}, given by 
\begin{equation}
\dot{n}_{\chi}+3Hn_{\chi}=2\left(1-\frac{n_{\chi}^2}{(n^{\rm eq}_{\chi})^2}\right) R_\chi,
\label{Eq:Boltz}
\end{equation}
where 
 \beq
 R_\chi=R^{\bar{f}f}_\chi+R^{Z'}_\chi + R^{Z'Z'}_\chi \, , 
 \label{Eq:rateSum}
 \eeq
and the factor of 2 accounts for the fact that each process produces or annihilates two DM particles\footnote{Note that here we have assumed that $Z'$ tracks its 
equilibrium number density. This is reasonable, because the 
dominant process which will be responsible for keeping $Z'$ in equilibrium is the inverse decay process $\bar{f}f\rightarrow Z'$ which is very efficient. As a result, the $Z'$ bosons will reach thermal 
equilibrium with the SM bath and maintain equilibrium until their equilibrium number density becomes Boltzmann suppressed when $T\lesssim M_{Z'}$.}. This can be re-written using the scale factor $a$ as the dynamical variable rather than time, and replacing the number densities with $Y_i=n_i a^3$
to obtain 
\begin{equation}
\frac{dY_{\chi}}{da}=\frac{2a^2}{H(a)}\left(1-\frac{Y_{\chi}^2}{(Y^{\rm eq}_{\chi})^2}\right) R_\chi \, ,
\label{Eq:Boltz2}
\end{equation}
where each of the rates $R$ and $H$ are functions of $a$.

Importantly, because we are studying ultra-relativistic freeze-out in addition to WIMP-like freeze-out and freeze-in, 
we cannot suffice with Boltzmann statistics. Instead, we must use the full quantum distribution functions for $f_{i}(E_{i},T)$ in Eq.~(\ref{Eq:rate}). We then should be clear about the limits of integration in Eq.~(\ref{Eq:rate}). Consider 
for example the $\bar{f}f\rightarrow \bar{\chi}\chi$ process. 
We recall that $E_{1}$ and $E_{2}$ refer to the energies of 
the initial state particles, which in this case are SM 
fermions, $f(E_{i},T)$. Because we neglect the SM fermion 
masses, the naive domains of integration would be $[0,\infty)$ for $E_{1}$ and $E_{2}$, and $[-1,1]$ for $\cos(\theta_{12})$. However, with these limits, we would 
frequently encounter $s<4m_{\chi}^2$, which would violate 
conservation of energy. 
In other words, for a given value of $E_1$, the integration bound of $E_2$ must satisfy the constraint 
$s=2E_1E_2(1-\cos \theta_{12})>4m_\chi^2$, bearing in mind that $E_1>\epsilon$ cannot be arbitrarily small since the fermions are considered relativistic.
As a result, we instead use the following domains of integration: 
\begin{gather}
\epsilon < E_1<\infty~\,,
\nonumber
\\
\frac{m_\chi^2}{E_1} < E_2 < \infty\,,
\nonumber
\\
-1 < \cos(\theta_{12})<1-2\frac{m_\chi^2}{E_1E_1}\,,
\end{gather}
where we choose $\epsilon$ sufficiently small such that the final result is insensitive to $\epsilon$.

\subsubsection{Freeze-in and UFO}

We can now derive analytic estimates of the relic abundance of DM when 
it is produced during the reheating era either by FI or UFO. Recall that the reheating era begins when inflation ends, and the inflaton begins 
oscillating about the minimum of its potential.  We consider the possibility 
that DM is produced from $a_{\rm end}$ to $a_{\rm RH}$,  where $a_{\rm end}$ is the scale factor at the beginning of reheating or, equivalently,
at the end of inflation 
and $a_{\rm RH}$ is the scale factor when the reheating ends. 
For FI production, the DM 
number density is always significantly lower than the equilibrium number 
density, such that the creation term in the Boltzmann equation is relevant 
while the annihilation term is negligible. 
This corresponds to neglecting the term proportional to $Y_\chi^2$ in Eq.~(\ref{Eq:Boltz2}).
On the other hand, at larger coupling or smaller mediator mass, the DM may 
come into equilibrium with the SM bath. This enables DM freeze-out (either WIMP-like or UFO) to occur during reheating or after reheating has completed.
When UFO occurs while the Universe is still in the reheating
phase, the DM falls out of equilibrium but SM bath particles are still efficiently produced by inflaton decay. 
This leads to the equilibrium number density in the SM bath
increasing relative to the actual DM number density.
As a result, the annihilation term in the Boltzmann equation quickly becomes negligible after UFO, similar to freeze-in. On the other hand, the initial condition for UFO is distinct from the initial condition for freeze-in. 
Namely, for freeze-in, the initial DM 
abundance is assumed to be zero ($Y_{\chi}(a_{\rm end})=0$). 
However, this hypothesis can be questioned, since gravitational interactions are also capable of producing DM during \cite{Garcia:2025rut} and after inflation \cite{Barman:2022qgt}.
In contrast, for UFO, {\it there 
is an initial abundance} when the DM freezes out ultra-relativistically, 
given by $Y_{\rm FO}=Y_{\chi}(a_{\rm FO})$.
This is always the case for DM that has experienced a period of thermal equilibrium before the freeze-out.
We stress that the cases for WIMPs and UFO DM  have the advantage not relying on non-thermal initial conditions.
Apart from the initial 
condition and the limits of integration, the evolution is the same for FI and UFO.

Neglecting the annihilation term, the Boltzmann equation (\ref{Eq:Boltz2}) becomes 
\beq
\frac{dY_{\chi}}{da}=\frac{2a^2 R_{\chi}(a)}{H(a)}\,,
\label{Eq:BoltzFIUFO}
\eeq
 where the Hubble parameter during post-inflationary reheating  is given by Eq.~(\ref{Eq:Hubble}).
 For an analysis of FI and UFO 
 under general reheating scenarios beyond the case of a quadratic minimum, see \cite{Henrich:2024rux,Henrich:2025gsd}.  The DM production rate per 
 unit volume, $R_{\chi}$, is computed from Eq.~(\ref{Eq:rate}), 
 summing over all initial states. The 
 integration leads to the rate as a function of temperature, $R_{\chi}(T)$. 
 In general, this will be a numerical result rather than a simple function 
 of $T$. However, it is possible to obtain a good analytical 
 approximation for relevant rates, 
 such as $\bar{f}f\rightarrow \bar{\chi}\chi$. 
 In the regime $m_{\chi}\ll T \ll M_{Z'}$,
 Eqs.~(\ref{Eq:rateVV}-\ref{Eq:rateAA}) give $R_\chi \propto \frac{T^8}{M_{Z'}^4}$ ($n=2$), whereas for $T \gg M_{Z'},m_{\chi}$, Eq.~(\ref{Eq:rateHighTemp}) gives
  $R_\chi\propto T^4$ ($n=-2$).
 We can then express $R_{\chi}(a)$ as a function of 
 the scale factor, using the relationship between the SM bath temperature 
 and the scale factor $a$ during reheating, determined by 
 Eqs~(\ref{Eq:rhoRad}) and (\ref{Eq:Hubble}). In each of the temperature 
 regimes mentioned above, the DM is relativistic such that the number 
 density is of the form $n^{\rm eq}_{\chi}\propto T^3$. Recalling that $R_{\chi}=\langle \sigma v \rangle n_{\rm eq}^2$, we find that $R_{\chi}$ 
 can be closely approximated by a simple polynomial in either $T$ or $a$. This enables us 
 to integrate Eq.~(\ref{Eq:BoltzFIUFO}) with respect to $a$.

 For the limits 
 of integration, we will use $a_{\rm end}$ as the lower limit and either $a_{\rm m}$ (defined as the scale factor when $T = m_{\chi}$) 
 for $m_{\chi} > T_{\rm RH}$ or $a_{\rm RH}$ for $m_{\chi} < T_{\rm RH}$ as the upper 
 limit.\footnote{In all of our numerical results we actually integrate well past $a_{\rm m}$ for the $m_\chi>\Trh$ case to ensure we account for any additional production when the rate is Boltzmann suppressed. However, taking $a_m$ as the upper limit of integration is useful here for our analytic approximations.} We anticipate two distinct results for the analytic 
 estimate of the final abundance depending on whether the interaction is UV 
 or IR dominated. 
 For UV dominant interactions, most of the DM is produced near $T=T_{\rm max}$. \footnote{For interactions mediated by a $Z'$ gauge boson, when $T> M_{Z'}$, $n=-2$, corresponding to IR production. However, when $T<M_{Z'}$, we transition to $n=2$ which may correspond to UV production depending on the details of reheating and hence most of the DM production will occur near $T = M_{Z'}$ rather than $\Tmax$. Alternatively, if the interaction reaches equilibrium and the DM undergoes UFO, the UV scale corresponds to $T_{\rm FO}$ described below. } 
 For an IR interaction, the production occurs primarily 
 at $m_{\chi}$ or $T_{\rm RH}$. 
We see therefore that the hierarchy between $m_\chi$, $M_{Z'}$ 
 and $\Trh$ will influence both the integration limits and the form of the rate $R_\chi(a)$. 
 When $R^{\bar{f}f}_\chi \gg R^{Z'}_\chi,  R^{Z'Z'}_\chi$, so that $R_\chi \simeq R^{\bar{f}f}_\chi$ and when $M_{Z'} \gg \Trh, m_\chi$, we can approximate
 \beq
R_\chi  \simeq n_{\rm eq}^2 \times \beta \frac{T^2}{M_{Z'}^4} \,, 
\label{Eq:rategeneric}
\eeq
corresponding to $n=2$. 
We can now distinguish four cases:

\vskip 0.5cm

\noindent 
(i) For UFO with $m_{\chi} < T_{\rm RH}<M_{Z'}$, 
\beq
Y_\chi(a_{\rm RH}) \approx Y_{\infty}= Y_{\rm FO} + 
\frac{9}{4 \sqrt{3}}\frac{\beta 
}{\sqrt{\alpha}}
\frac{g_\chi^2\zeta(3)^2}{\pi^4} 
\frac{T_{\rm RH}^{6}M_{P}}{M_{Z'}^{4}}\arh^3 \left[
1- \left(\frac{\aFO}{\arh}\right)^{\frac{3}{2}}\right]\,,
\label{Eq:yrhUFOmLeqTRH}
\eeq

\noindent (ii) For UFO with $T_{\rm RH}<m_{\chi}<M_{Z'}$,
\beq
Y_\chi(\am) \approx Y_\chi(a_{\rm RH}) \approx Y_{\infty}= Y_{\rm FO} + 
\frac{9}{4 \sqrt{3}}\frac{\beta 
}{\sqrt{\alpha}}
\frac{g_\chi^2\zeta(3)^2}{\pi^4} 
\frac{T_{\rm RH}^{6}M_{P}}{M_{Z'}^{4}}\arh^\frac32 \left[
\am^\frac32- \aFO^\frac32\right]\,,
\label{Eq:yrhUFOmGeqTRH}
\eeq

\noindent 
(iii) For freeze-in with $m_{\chi}<T_{\rm RH}<M_{Z'}$, 
\beq
Y_\chi(a_{\rm RH}) \approx Y_{\infty}= 
\frac{9}{4 \sqrt{3}}\frac{\beta 
}{\sqrt{\alpha}}
\frac{g_\chi^2\zeta(3)^2}{\pi^4} 
\frac{T_{\rm RH}^{6}M_{P}}{M_{Z'}^{4}}\arh^3 \left[
1- \left(\frac{\aend}{\arh}\right)^{\frac{3}{2}}\right]\,,
\label{Eq:yrhFImLeqTRH}
\eeq

\noindent 
(iv) For freeze-in with $T_{\rm RH}<m_{\chi}<M_{Z'}$,
\beq
Y_\chi(\am) \approx Y_\chi(a_{\rm RH}) \approx Y_{\infty}=
\frac{9}{4 \sqrt{3}}\frac{\beta 
}{\sqrt{\alpha}}
\frac{g_\chi^2\zeta(3)^2}{\pi^4} 
\frac{T_{\rm RH}^{6}M_{P}}{M_{Z'}^{4}}\arh^\frac32 \left[
\am^\frac32- \aend^\frac32\right]\,,
\label{Eq:yrhFImGeqTRH}
\eeq
where we used Eq.(\ref{Eq:neq}) for $n_{\rm eq}$ and
\beq
\beta=\frac{49 \pi^3 \Sigma_{\rm tot}}{259200}\left(\frac{4\pi^2}{3g_\chi \zeta(3)}\right)^2=\frac{49 \pi^7 \Sigma_{\rm tot}}{145800 g_\chi^2 \zeta(3)^2} \,,
\eeq 
with $\Sigma_{\rm tot}$ given by Eq.~(\ref{Eq:sigmatot}).  
Notice that while these expressions have a similar form, particularly with respect to the out-of-equilibrium production term on the RHS, the limits of integration and initial conditions are distinct in each case, which leads to important phenomenological differences between UFO and FI, such as restrictions on the allowed couplings, mediator mass, and DM mass. 
Recalling that the number density $n_\chi = Y/a^3$, the determination of the relic abundance
requires ratios of the scale factors involving $\aend, \arh, a_{\rm m}$ and $\aFO$, in terms of $\rho_{\rm end}, \rho_{\rm RH}, m_\chi$, and $\TFO$ (evaluated below). For $T>\Trh$ ($a<\arh$), and $m_\chi>\Trh$ ($a_{\rm m} < \arh$), we have
\beq
\left(\frac{\aend}{\arh} \right)^3 = \frac{\rho_{\rm RH}}{\rho_{\rm end}} \; ;  \qquad
\left(\frac{\aFO}{\arh} \right)^3 = \left( \frac{\Trh}{\TFO} \right)^8 \; ; \qquad \left(\frac{a_{\rm m}}{\arh} \right)^3 = \left( \frac{\Trh}{m_\chi} \right)^8 \, , 
\label{arats1}
\eeq 
if not for For $T<\Trh$ ($a<\arh$), and $m_\chi<\Trh$ ($a_{\rm m} < \arh$)
\beq
\left(\frac{\aFO}{\arh} \right) = \left( \frac{\Trh}{\TFO} \right) \; ; \qquad \left(\frac{a_{\rm m}}{\arh} \right) = \left( \frac{\Trh}{m_\chi} \right) \, . 
\eeq

The value of $Y_{\rm FO}$ is the same for cases (i) and (ii) and can be determined as follows \cite{Henrich:2025gsd}, 
\begin{align}
Y_{\rm FO} = Y_{\chi}(a_{\rm FO}) = n_{\chi}(a_{\rm FO})a_{\rm FO}^3 
 = & { \frac34}\frac{g_{\chi} \zeta(3)}{\pi^2}T_{\rm FO}^3 a_{\rm RH}^3 \left(\frac{T_{\rm RH}}{T_{\rm FO}}\right)^{8}  \nonumber \\
= & { \frac34}\frac{g_{\chi} \zeta(3)}{\pi^2}\left[\sqrt{\frac{\alpha}{3}}\frac{2\pi^2 M_{Z'}^{4}}{g_{\chi} \zeta(3) \beta  M_P }\right]^{-5} T_{\rm RH}^{18} a_{\rm RH}^3\, ,
 \label{Eq:yfo}
\end{align}
where $\TFO$ is obtained from the freeze-out condition during matter domination assuming $\Trh < T_{\rm FO}$
\beq
\frac{R_\chi}{n_{\rm eq}} \left. \right|_{\TFO} = \frac32 H(\TFO)\,,
\label{freezeout}
\eeq
which gives
\beq
\TFO^{>\Trh}=\sqrt{\frac{\alpha}{3}}\frac{2 \pi^2}{g_\chi \zeta(3)\beta 
}\frac{M_{Z'}^4}{\Trh^2 M_P}\simeq 290~{\rm GeV}~\left(\frac{M_{Z'}}{10^6~\rm GeV}\right)^4\left(\frac{100~\rm GeV}{\Trh}\right)^2\,,
\label{Eq:TFO}
\eeq 
where we have used $H(\TFO) = \sqrt{\alpha/3}\TFO^4/(\Trh^2 M_P)$ from Eq.~(\ref{Eq:Hubble}). 
In the event that $\TFO<\Trh$, freeze-out
occurs during the radiation era and the factor of $\frac32$ becomes 2 in the right hand side of Eq.~(\ref{freezeout}) with $H(\TFO) = \sqrt{\alpha/3}\TFO^2/ M_P$.
In this case, we have
\beq
\TFO^{<\Trh} = \left( \sqrt{\frac{\alpha}{3}}\frac{8 \pi^2}{3 g_\chi \zeta(3)\beta 
}\frac{M_{Z'}^4}{ M_P} \right)^\frac13 \simeq 160~{\rm GeV} 
\left(\frac{M_{Z'}}{10^6~\rm GeV}\right)^\frac43 \, .
\label{Eq:TFObis}
\eeq 
If one takes $M_{Z'} = 100$~GeV, we would find $\TFO \sim 1$~MeV similar to the freeze-out temperature for SM model neutrinos. 

Note that in Eq.~(\ref{Eq:yrhUFOmLeqTRH}), the first term ($\propto \Trh^{18}$) is negligible compared to the 2nd term since production is IR dominated and the final abundances in cases (i) and (iii) are nearly identical. 
The condition on $M_{Z'}$ should, naively, be $M_{Z'}> T_{\rm max}$ in order to use the approximation $\langle \sigma v \rangle \sim \frac{T^2}{M_{Z'}^4}$ during the entire reheating process. However, since production is IR-dominated, the condition $M_{Z'}>\Trh,m_\chi$ is sufficient. Note that in this work, we consider only $M_{Z'}>T_{\rm RH}$.

It is worth noting the similarities between (i) and (iii) as well as between (ii) and (iv).  The main differences are the limits of integration and initial conditions. While in the FIMP case, $n_\chi(\aend)$ is assumed to be zero, while in UFO, $n_\chi$ is {\it fixed} by the conditions of thermal equilibrium.  UFOs are thus more predictable and potentially more observable than FIMPs. 
It is also worth noting that when the DM production is IR dominated, $Y_{\rm FO}$ is much less than the out-of-equilibrium contribution post-freeze-out in Eqs.~(\ref{Eq:yrhUFOmLeqTRH}) and (\ref{Eq:yrhUFOmGeqTRH}). In these cases, the \textit{form} of the final relic abundance for FIMPs and UFO DM are the same. This might lead one to ask: how is UFO different from freeze-in? The differences are many. In addition to the fact that UFO DM enters equilibrium and thus the initial conditions are known and well-motivated, the FIMP and UFO mechanisms are associated with vastly different regions of parameter space. Indeed, FIMPs and UFO DM are compatible with non-overlapping regions of the large, multi-dimensional parameter space spanned by $T_{\rm RH}, M_{Z'}, m_{\chi}$, and the relevant couplings; and the correct final abundance will only be obtained by one of the relevant mechanisms. We will focus on these differences in the results section below. As a general rule, stronger couplings, smaller mediator mass, and higher reheating temperatures will be more likely to be compatible with UFOs rather than FIMPs \cite{HMO2}. 

In previous work pertaining to WIMPs and FIMPs, the UFO parameter space has almost always been neglected for several reasons. In WIMP-related studies, one typically restricts the parameter space such that $x_{\rm FO}=m_{\chi}/T_{\rm FO}\gtrsim3$, in order to avoid hot relic DM. However, as we have previously demonstrated, DM which freezes out for $x_{\rm FO}<1$ does not in fact lead to hot relic DM when freeze-out occurs during reheating because there is ample time for the momentum to redshift during and after reheating, prior to structure formation \cite{Henrich:2025gsd}. Similarly, the UFO regime has been neglected in freeze-in studies because for sufficiently strong interactions, the DM equilibrates and therefore is no longer produced by standard freeze-in. As a result, UFO DM is truly distinct from both FIMPs and WIMPs. Moreover, even the similarity in the form of the final abundance between FIMPs and UFO DM disappears when one considers the case of UV production, in which case the UFO mechanism is even more distinctive \cite{Henrich:2025gsd}. 

Before concluding this subsection, we emphasize that in contrast to post-reheating UFO, even light DM candidates produced via UFO during reheating with $m_\chi \lesssim 1$~MeV
do not create problems with either the big bang nucleosynthesis (BBN) constraint on the number of light relativistic degrees of freedom, or the bounds from wiping out structure on small cosmological distance scales. In \cite{Henrich:2025gsd}, we showed that the bound on $\Delta N_{\rm eff} < 0.18$ from BBN \cite{ysof} is easily satisfied, so long as $\Trh<\TFO$, as is assumed here. More precisely for a Dirac DM candidate, it is sufficient that $\Trh < 0.7 (g_{\rm FO}^2/(g_{\rm BBN} g_{\rm RH}))^\frac15 \TFO \simeq 1.1 \TFO$
with $g_{\rm FO} =  g_{\rm RH} = 427/4$ and $g_{\rm BBN} = 43/4$. This is satisfied in all the models considered here. There is also a constraint from structure formation, namely $m_\chi > 5$~keV  \cite{Henrich:2025gsd}.

\subsubsection{WIMP-like freeze-out}

In contrast to UFO DM candidates, which freeze-out relativistically during reheating, or FIMPs which never freeze-out, as they were never in equilibrium, WIMP-like freeze-out occurs when the DM candidate becomes non-relativistic. 
We next consider the case of WIMP-like freeze-out which may occur during reheating. 
The procedure is quite similar to the 
vanilla WIMP case \cite{Gondolo:1990dk}, except that the thermal bath does not satisfy the isentropic condition, $T\propto 1/a$.
To estimate the final abundance when the DM undergoes WIMP-like freeze-out 
during reheating, we first need an estimate of the freeze-out temperature. 
To do this, we simply relate the DM production rate to the Hubble parameter as in Eq.~(\ref{freezeout}) remembering to take the non-relativistic limit.
 When WIMP-like FO occurs, the equilibrium number density of DM 
becomes Boltzmann suppressed. 
When $m_{\chi} \lesssim M_{Z'}$, the non-relativistic limit of the thermally averaged cross section becomes $R_\chi/n^2_{\rm eq} \approx {\tilde \beta} \frac{m_{\chi}^2}{M_{Z'}^4}$.
Then, from the Hubble parameter, Eq.~(\ref{Eq:Hubble}), we have
\begin{equation}
\tilde \beta\frac{m_{\chi}^2}{M_{Z'}^4}\frac{g_{\chi}}{(2\pi)^{3/2}}(m_{\chi}T_{\rm FO})^{3/2}e^{-m_{\chi}/T_{\rm FO}}=\frac32 \sqrt{\frac{\alpha}{3}}\frac{T_{\rm FO}^4}{T_{\rm RH}^2 M_{P}}
\,.
\label{findxfo}
\end{equation}
where 
\beq
\tilde{\beta}=\frac{49 \pi^3 \Sigma_{\rm tot}}{259200}\left(\frac{(2\pi)^{3/2}}{g_\chi}\right)^2=\frac{49 \pi^6 \Sigma_{\rm tot}}{32400 g_\chi^2} \,.
\eeq
We can then change variables to $x_{\rm FO}=m_{\chi}/T_{\rm FO}$ which gives
\begin{equation}
x_{\rm FO}^{\frac{5}{2}}e^{-x_{\rm FO}}=
\frac{\sqrt{6 \pi^3 \alpha}}{g_\chi \tilde \beta}\frac{M_{Z'}^4}{m_\chi \Trh^2 M_P}\,,
\label{Eq:xFOreheating}
\end{equation}
with a solution for $x_{\rm FO}$ given by
\begin{equation}
x_{\rm FO}=-\frac{5}{2}\mathcal{W}_{-1}\left(-\frac{2}{5}\left(\frac{\sqrt{6\pi^3\alpha}}{g_{\chi}  \tilde\beta}\frac{M_{Z'}^4}{m_{\chi}T_{\rm RH}^2 M_{P}}\right)^{2/5}\right)
\label{Eq:xFO4}
\end{equation}
where $\mathcal{W}_{-1}$ denotes the -1 branch of the Lambert function. We note that, unsurprisingly, this result is sensitive to $M_{Z'}$, $T_{\rm RH}$, and $m_{\chi}$. 

For $m_\chi > M_{Z'}$, the production rate is altered with $R_\chi/n^2_{\rm eq} \approx \frac{\hat{\beta}}{m_{\chi}^2}$ and the left hand side of Eq.~(\ref{findxfo}) is altered and we have
\beq
\frac{\hat{\beta}}{m_{\chi}^2} \frac{g_{\chi}}{(2\pi)^{3/2}}(m_{\chi}T_{\rm FO})^{3/2}e^{-m_{\chi}/T_{\rm FO}}=\frac32 \sqrt{\frac{\alpha}{3}}\frac{T_{\rm FO}^4}{T_{\rm RH}^2 M_{P}}\,,
\label{Eq:xFOreheatingbis}
\eeq
where
\beq
\hat{\beta}=\frac{\Sigma_{\rm tot}\pi^2}{1728 g_\chi^2}\,.
\label{Eq:betahat}
\eeq
Once again, we can replace $\TFO$ with $m_\chi/\xfo$
to obtain
\begin{equation}
x_{\rm FO}^{\frac{5}{2}}e^{-x_{\rm FO}}=
\frac{\sqrt{6 \pi^3 \alpha}}{g_\chi \hat \beta}\frac{m_\chi^3}{\Trh^2 M_P}\,,
\label{Eq:xFOreheating3}
\end{equation}
leading to a new solution for $x_{\rm FO}$
\beq
x_{\rm FO} = 
-\frac{5}{2}\mathcal{W}_{-1}\left(-\frac{2}{5}\left( \frac{\sqrt{6 \pi^3 \alpha}}{g_{\chi} \hat{\beta}}\frac{m_{\chi}^3}{T_{\rm RH}^2 M_{P}}\right)^{2/5}\right) \, .
\label{xfo3}
\eeq
Note that this equation, which determines the FO temperature, is quite different from the one found in the conventional WIMP literature \cite{Gondolo:1990dk,Arcadi:2017kky}. Indeed, in the case of WIMP-type
decoupling {\it during the reheating period}, there is an explicit dependence of $x_{\rm FO}$ on $\Trh$, since additional entropy from inflaton decay is produced between $\TFO$ and $\Trh$. 
For example, for $(m_\chi,\Trh,M_{Z'})=(10^3, 0.1, 10^4)$ GeV, one obtains
$x_{\rm FO} \simeq 13$, 
whereas for $(m_\chi,\Trh,M_{Z'})=(10^3, 10, 10^4)$ GeV, we have $x_{\rm FO} \simeq 23$.
Since the non-relativistic number density in Eq.~(\ref{Eq:neq}) at $\TFO$ depends exponentially on $x_{\rm FO}$, the dependence of $x_{\rm FO}$ on $\Trh$ can drastically affect the relic abundance.

In comparison, for the case of a universe dominated by radiation, with $\TFO < \Trh$, we can obtain the analogous expression for $x_{\rm FO}$.  Indeed, for WIMP-like FO during radiation domination (the standard WIMP FO process), the freeze-out condition $\Gamma=2 H$ 
gives us 
\begin{equation}
x_{\rm FO}^{1/2}e^{-x_{\rm FO}}=2 \sqrt{\frac{\alpha}{3}}\frac{(2 \pi)^{3/2}}{g_{\chi} \tilde{\beta}}\frac{M_{Z'}^4}{m_{\chi}^3 M_{P}},
\end{equation}
which is independent of $\Trh$ and should be compared with Eq.~(\ref{Eq:xFOreheating}).
We then find
\beq
x_{\rm FO}=-\frac{1}{2}\mathcal{W}_{-1}\left(-\frac{64 \pi^3\alpha}{3 g_\chi^2 \tilde{\beta}^2}\frac{M_{Z'}^8}{m_\chi^6 M_{P}^2}\right) \, .
\eeq
For example with
$(m_\chi,\Trh,M_{Z'})=(10^3, >50, 10^4)$ GeV
$x_{\rm FO}\simeq 26$.

While Eqs.~(\ref{Eq:xFO4})and (\ref{xfo3}) provide analytic estimates of $x_{\rm FO}$ for WIMP-like freeze-out during reheating for the $Z'$ portal interaction in two distinct limits, it is also possible to find the precise value of $x_{\rm FO}$ corresponding to the boundary between UFO and WIMP-like freeze-out in a straightforward manner. In particular, we know that the quantity $Y_{\rm eq}(a)=n_{\rm eq}a^3$ initially rises during the reheating period when the DM is ultra-relativistic ($n_{\rm eq}\propto T^3=T_{\rm RH}^3\left(\frac{a_{\rm RH}}{a}\right)^{9/8} \Rightarrow Y_{\rm eq}\propto a^{15/8}$). However, if $m_\chi>T_{\rm RH}$, the equilibrium DM number density will become Boltzmann suppressed ($n_{\rm eq}\propto (m_\chi T)^{3/2}e^{-m_\chi/T}$) for sufficiently low $T$ or sufficiently high $a<a_{\rm RH}$. As a result, $Y_{\rm eq}$ will eventually also become Boltzmann suppressed. Therefore, it is possible to find the maximum of $Y_{\rm eq}$ by simply setting $\frac{dY_{\rm eq}}{da}=0$. The result will provide the scale factor at which $Y_{\rm eq}$ is maximized, which is also the natural boundary between WIMP-like FO and UFO. We find the simple result
\beq
\frac{dY_{\rm eq}}{da}=0 \hspace{2mm} \Rightarrow \hspace{2mm} x_{\rm FO}|_{UFO/WIMP}=13/2 \, .
\label{Eq:xFOboundary}
\eeq
This is consistent with what was previously reported in the first thorough study of WIMPs during reheating \cite{Bernal:2022wimp}, where the authors found $x_{\rm FO}\approx 6$ as their approximate lower limit. Strictly speaking, the range $1/3 \lesssim x_{\rm FO}< 13/2$ may more properly be called semi-relativistic or relativistic freeze-out, rather than ultrarelativistic, while $x_{\rm FO}\lesssim 1/3$ may be called ultrarelativistic freeze-out. For simplicity, we will categorize relativistic and ultrarelativistic freeze-out together (all $x_{\rm FO}<13/2$) as UFO. Indeed, for most of the UFO parameter space reported in our results section, $x_{\rm FO}\ll 1/3$ and is therefore aptly categorized as UFO.

\section{Results}
\label{results}

\begin{figure}[!ht]
\centering
\vskip .2in
\includegraphics[width=\textwidth]{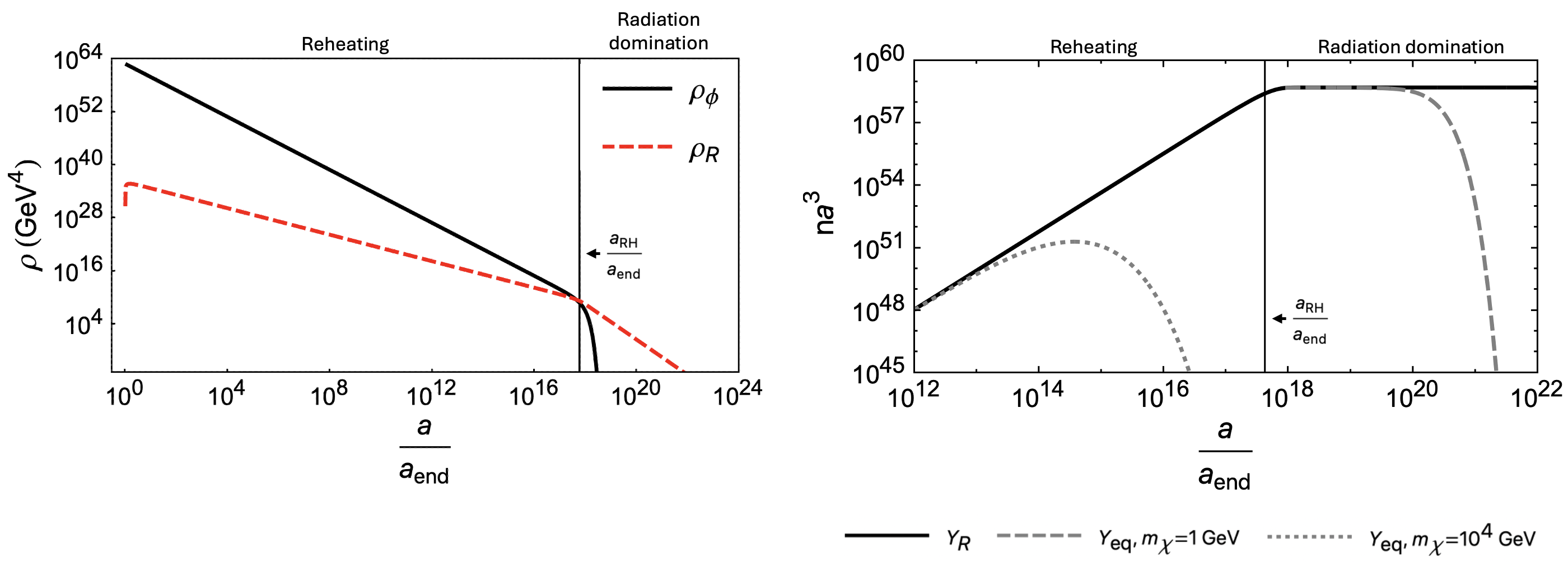}
\caption{\em \small Evolution of the energy densities of the oscillating inflaton condensate and the SM radiation bath during and after the reheating period (left panel). Evolution of the co-moving number of SM radiation and the co-moving DM equilibrium number densities (right panel) for two choices of DM mass, $m_\chi=1$~GeV ($m_\chi<T_{\rm RH}$) and $m_\chi=10^4$~GeV ($m_\chi>T_{\rm RH}$). $T_{\rm RH}=100$~GeV in both panels. The co-moving SM radiation number density, $Y_{R}$, in the right panel has been normalized to that of a representative fermionic SM bath species (e.g. electron) for direct comparison with fermionic DM.}
\label{Fig:ReheatingBasics}
\end{figure}

We first illustrate some of the basic features of our treatment of the non-instantaneous reheating period. In the left panel of Fig.~\ref{Fig:ReheatingBasics}, the evolution of the energy density of the oscillating inflaton condensate and the SM radiation bath are depicted as a function of the scale factor, $a$, normalized to the scale factor at the end of inflation, $a_{\rm end}$. At the beginning of the reheating period, the inflaton condensate (black) stores an initial energy density corresponding to its energy density at the end of inflation $\rho_{\phi}(a_{\rm end})\approx 10^{62} \text{ GeV}^4$. The oscillating condensate evolves throughout the reheating period as $\rho_{\phi}\propto a^{-3}$. The energy density of the SM radiation bath (red) quickly reaches a max temperature $T_{\rm max}>T_{\rm RH}$ before evolving as $\rho_{R}\propto a^{-3/2}$, or equivalently $T\propto a^{-3/8}$. When $a\approx a_{\rm RH}$, the SM radiation energy density overtakes the oscillating inflaton's energy density, and we recover the ordinary radiation dominated era. In the right panel of Fig.~\ref{Fig:ReheatingBasics}, we depict the co-moving number of SM radiation ($Y_{R}$) along with the co-moving DM equilibrium number density ($Y^{\rm eq}_{\chi}$) for two choices of DM mass. During the reheating era, the co-moving radiation number evolves as $Y_{R}\propto T^3a^3 \propto a^{15/8}$. The increase in $Y_{R}$ during reheating is due to continuous inflaton decay to SM particles. In contrast, during the standard radiation dominated era which follows reheating, the co-moving radiation number is constant, up to changes in the number of relativistic degrees of freedom. The equilibrium number density of DM tracks the SM radiation number density until the SM bath temperature drops below $T\approx m_\chi$, when its equilibrium number density becomes Boltzmann suppressed. For $T_{\rm RH}=100$~GeV and $m_\chi=10^4$~GeV, this Boltzmann suppression of the DM equilibrium number density happens during the reheating era (dotted gray line in right panel of Fig.~\ref{Fig:ReheatingBasics}). In contrast, for $m_\chi=1$~GeV, the equilibrium number density becomes Boltzmann suppressed during the radiation dominated era (dashed gray line), once reheating has already completed. These examples of the evolution of the equilibrium DM number density are critical for understanding how DM production occurs during vs. after reheating, for WIMP-like FO, UFO, and FI. 

\subsection{WIMPs, FIMPs, and UFOs}

\begin{figure}[!ht]
\centering
\vskip .2in
\includegraphics[width=0.75\textwidth]{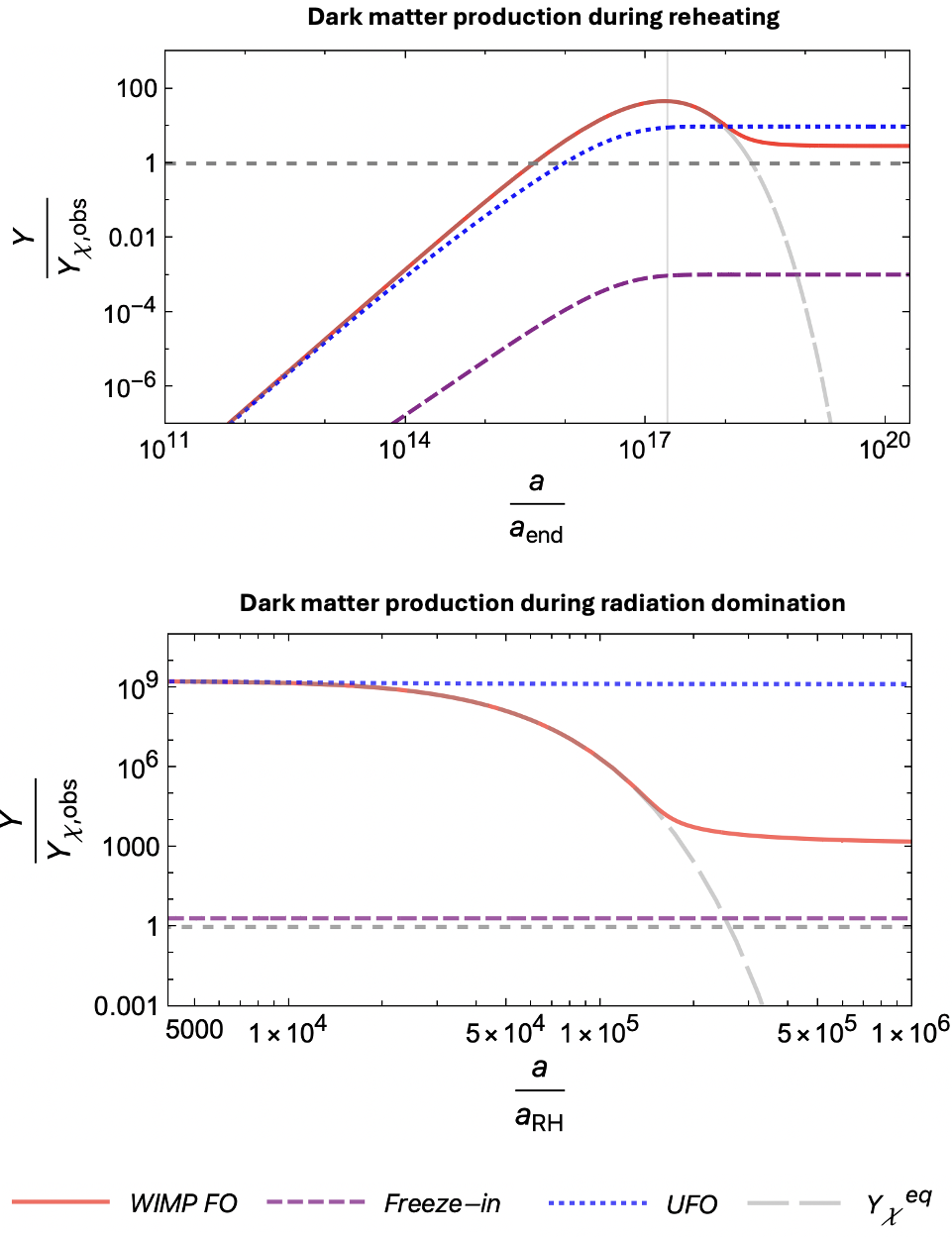}
\caption{\em \small The evolution of the co-moving DM number density $Y_\chi$ for WIMP-like FO (solid red), UFO (dotted blue), and FI (dashed purple) when DM production occurs during reheating (top panel, focus of this study) or during radiation domination (bottom panel, standard scenario). Parameter choices are $M_{Z'}=10$~TeV, $m_{\chi}=100$~GeV for both panels. For the top panel, $T_{\rm RH}=1$~GeV and the couplings are $V_\chi=1$ (red), $V_\chi=0.1$ (blue), and $V_\chi=0.01$ (purple). For the bottom panel, $T_{\rm RH}=10^{10}$~GeV and the couplings are $V_\chi=1$ (red), $V_\chi=0.01$ (blue), and $V_\chi=10^{-5}$ (purple). The horizontal gray dashed lines indicate the correct final abundance. The thin vertical line in the top panel corresponds to $T=\frac{2}{13}m_\chi$, which is the natural boundary between non-relativistic and relativistic FO during reheating.}
\label{Fig:WIMPFIMPUFO}
\end{figure}

Next, we illustrate and compare the three different DM production mechanisms: WIMP-like (non-relativistic) freeze-out, UFO, and FI in Fig.~\ref{Fig:WIMPFIMPUFO}. We depict the evolution of the co-moving DM abundance, $Y_\chi=n_\chi a^3$, for all three mechanisms when DM production occurs during reheating (top panel) which is the focus of this study, alongside the standard scenario where DM production occurs during radiation domination (bottom panel). For DM production during reheating, we chose three different values of the vector coupling $V_\chi=1$ (red), $V_\chi=0.1$ (blue), and $V_\chi=0.01$ (purple) corresponding to WIMP-like FO, UFO, and FI respectively; and we depict the evolution of $Y_{\chi}(a)$ for each along with the co-moving equilibrium DM number density, $Y^{\rm eq}_{\chi}$. For the purposes of this comparison, we used $M_{Z'}=10$~TeV, $m_\chi=100$~GeV, $A_\chi=0$, and $T_{\rm RH}=1$~GeV. The quantity on the vertical axis is $Y_{\chi}/Y_{\chi, \text{obs}}$, where $Y_{\chi, \text{obs}}$ is the observed co-moving DM number density corresponding to $\Omega_\chi h^2=0.12$. 

In Fig.~\ref{Fig:WIMPFIMPUFO}, freeze-out occurs for the blue-dotted (UFO) and red-solid (WIMP) curves when they separate from the equilibrium DM number density (gray-dashed) curve. For UFO, this separation occurs while the DM is still ultrarelativistic ($T_{\rm FO}\gg m_\chi$) while for WIMP-like FO, the red curve departs from $Y^{\rm eq}_\chi$ when $T_{\rm FO}<m_\chi$. Recall that since we are plotting $Y$ as a function of the scale factor $a$, the temperature increases to the left.   It is easy to see how changing the coupling, $V_\chi$, can allow us to seamlessly shift between all three mechanisms. At strong coupling ($V_\chi=1$ in this case), the DM will remain in equilibrium for a long period of time due to rapid, efficient interactions with the SM bath, until $Y^{\rm eq}_\chi$ becomes Boltzmann suppressed and the DM will then undergo WIMP-like FO (red-solid). If we reduce the coupling (to $V_\chi=0.1$ in the top panel or $V_\chi=0.01$ in the bottom panel), we then transition to a regime where the interaction is strong enough for the DM to enter equilibrium, but \textit{not strong enough} to retain the DM in equilibrium all the way into the Boltzmann suppressed regime. In this case, the DM decouples from the SM radiation bath while the DM is still (ultra)relativistic (blue-dotted), prior to the regime of Boltzmann suppression. Finally, we can reduce the coupling even further ($V_\chi=0.01$ in the top panel or $V_\chi=10^{-5}$ in the bottom panel), such that the DM-SM interaction is sufficiently feeble and the DM no longer reaches equilibrium at any point in the universe's history and DM production proceeds via FI (purple-dashed). Thus, all three mechanisms (WIMP FO, UFO and FI) are seamlessly connected to one another; and we can transition from one to the next by varying either the coupling, as we show in Fig.~\ref{Fig:WIMPFIMPUFO}, or alternatively $M_{Z'}$ or $T_{\rm RH}$ (the latter only facilitates transitions between these mechanisms when production occurs during reheating). Note that while we elected to use a natural set of $V_\chi$ values for the purposes of illustration in Fig.~\ref{Fig:WIMPFIMPUFO} rather than tune $V_\chi$ to obtain the correct abundance ($Y_\chi/Y_{\chi, \text{obs}}=1$), it is not difficult to do this. Alternatively, $V_\chi$ can be held fixed and the correct abundance may be obtained by varying $M_{Z'}$, $m_\chi$, and/or $T_{\rm RH}$. Also note that the specific ranges of the coupling $V_\chi$ which lead to WIMP-like FO, UFO, or FI are generally different when FO occurs during reheating vs. radiation domination, which is why we chose different values of $V_\chi$ to illustrate UFO and FI in the top and bottom panels. The underlying reason for these differences is because the temperature dependence of the Hubble parameter, which determines $T_{\rm FO}$, is distinct during reheating ($H\propto T^4$) compared to radiation domination ($H\propto T^2$).

As previously mentioned, $Y^{\rm eq}_{\chi}$ (gray dashed line) increases throughout the reheating period (top panel) due to continuous production of SM radiation via inflaton decays, until $Y^{\rm eq}_\chi$ eventually becomes Boltzmann suppressed when $T\lesssim m_\chi$. This is in stark contrast to the analogous evolution of the co-moving DM equilibrium number density during radiation domination (gray dashed line, bottom panel), which remains constant prior to Boltzmann suppression. This is a critical distinction which makes the UFO mechanism much more interesting during reheating compared to radiation domination. In particular, the increase in $Y^{\rm eq}_\chi$ during reheating makes the blue dotted curve in the top panel of Fig.~\ref{Fig:WIMPFIMPUFO} non-trivial after freeze-out, while the blue curve in the bottom panel is trivial and immovable. If UFO occurs during radiation domination, the co-moving number $Y_\chi$ after freeze-out remains essentially constant (up to changes in the number of relativistic degrees of freedom), tracking $Y_R$ in Fig.~\ref{Fig:ReheatingBasics}. In contrast, as we see in the top panel of Fig.~\ref{Fig:WIMPFIMPUFO}, when UFO occurs during reheating, significant DM production can continue to occur \textit{after} freeze-out due to out-of-equilibrium interactions, leading to an increase in $Y_\chi$ by several orders of magnitude after freeze-out. In the case of WIMP-like FO (red) during reheating or during radiation domination, we see that post-freeze-out annihilations determine the final abundance, since FO occurs in the Boltzmann suppressed regime. This differs from both UFO and FI during reheating, where the out-of-equilibrium \textit{production} (not annihilation) determines the final abundance. The FI curve (purple) in the top panel shows the expected evolution of $Y_\chi$ due to continuous, out-of-equilibrium production. The FI curve in the bottom panel appears constant; this is because the production for this process is UV-dominated during radiation domination, with essentially all of the production happening very early. The vertical gray line in the top plot represents the natural boundary between non-relativistic FO and relativistic FO during reheating, corresponding to a bath temperature of $T=\frac{2}{13}m_\chi$ previously reported in Eq.~(\ref{Eq:xFOboundary}). From the top panel of Fig.~\ref{Fig:WIMPFIMPUFO}, we can see that if $T_{\rm FO}>\frac{2}{13}m_\chi$ (i.e. if freeze-out occurs to the left of the vertical gray line), the behavior will be similar to the blue curve, with out-of-equilibrium production primarily determining the final abundance. In contrast, if $T_{\rm FO}< \frac{2}{13}m_\chi$, non-relativistic FO will occur and the behavior will be similar to the red curve, with annihilations primarily determining the final abundance.

\subsection{Parameter space associated with the correct DM abundance}

\begin{figure}[!ht]
\centering
\vskip .2in
\includegraphics[width=\textwidth]{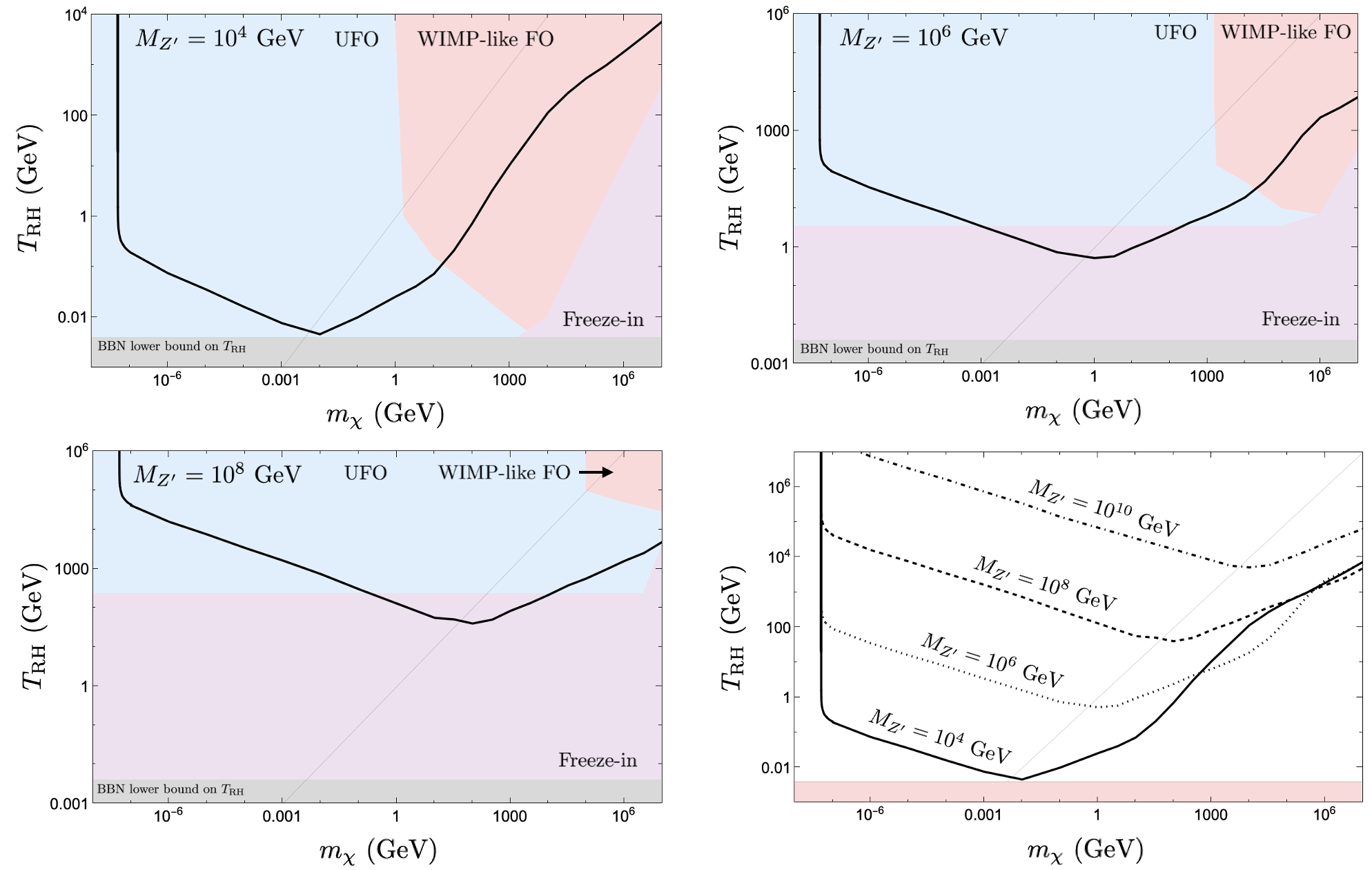}
\caption{\em \small The $(m_{\chi},T_{\rm RH})$ plane showing contours (black) for which $\Omega_{\chi}h^2=0.12$. Three choices of $M_{Z'}$ are displayed in separate figures illustrating the regions of parameter space associated with UFO (blue), WIMP-like FO (red), and freeze-in (purple) (top left: $M_{Z'}=10^4$~GeV, top right: $M_{Z'}=10^6$~GeV, bottom left: $M_{Z'}=10^8$~GeV).
In the bottom right panel, contours corresponding to the correct abundance are overlaid for four choices of $M_{Z'}$. Couplings were chosen to be $V_{f}=V_{\chi}=1$ and $A_{f}=A_{\chi}=0$. The thin gray lines correspond to $m_{\chi}=T_{\rm RH}$. The shaded region at the bottom of each plot is excluded due to the lower bound on $T_{\rm RH}$ from BBN ($T_{\rm RH}>4$~MeV). }
\label{Fig:PureVectorCoupling}
\end{figure}

Our main results are reported in Fig.~\ref{Fig:PureVectorCoupling}, 
where we show in the $(m_{\chi},T_{\rm RH})$ plane,
the parameter space for which the observed relic abundance $(\Omega_{\chi}h^2=0.12)$ is satisfied for different values of $M_{Z'}$ ($10^4$, $10^6$, $10^8$ and $10^{10}$~GeV). To obtain these plots, we solved Eq.~(\ref{Eq:rate}) numerically for fixed $M_{Z'}$ and $m_{\chi}$ to obtain 
$R_{\chi}(M_{Z'}, m_{\chi},T)$. 
We then numerically solved Eq.~(\ref{Eq:Boltz2}) for 
different values of $T_{\rm RH}$.  
To do this, we obtain $H(a)$ and each $R(a)$, which are distinct functions for each choice of $T_{\rm RH}$. Note that the integration limits also 
depend on $T_{\rm RH}$ through $\frac{a_{\rm RH}}{a_{\rm end}}$.
We then repeat this entire process for many choices of the DM mass 
$m_{\chi}$, while keeping $M_{Z'}$ fixed. The output of each of these 
calculations is the co-moving DM abundance $Y_{\chi}(a_{\rm RH})=n_{\chi}(a_{\rm RH})a_{\rm RH}^3$, which can easily be converted to a final relic abundance using \cite{mybook}
\begin{equation}
\Omega_{\chi}h^2\simeq5.88\times 10^6 \, \left(\frac{n_{\chi}(a_{\rm RH})}{T_{\rm RH}^3}\right)\left(\frac{m_\chi}{1~\rm GeV}\right)\,.
\label{Eq:omegageneric}
\end{equation}
We show the resulting isocontours with $\Omega_{\chi}h^2=0.12$ in Fig.~\ref{Fig:PureVectorCoupling}. 
Above each curve, for a given $M_{Z'}$, the DM is overabundant. Repeating this process for different 
choices of $M_{Z'}$, leads to multiple contours which can be overlaid in the same plane, shown in the bottom right panel of  Fig.~\ref{Fig:PureVectorCoupling}. 
We considered both $m_{\chi}>T_{\rm RH}$ and $m_{\chi}<T_{\rm RH}$, 
and placed a lower limit on $T_{\rm RH}>4$~MeV corresponding to the BBN bound \cite{tr4}.

We first consider a simple model where all SM fermions along with the DM fermion $\chi$ are charged under the new $U(1)'$ gauge symmetry, with equal vector couplings $V_{f}=V_\chi=1$ and axial vector couplings $A_{f}=A_\chi=0$. The sum 
over all the fermions in Eq.~(\ref{Eq:rate}) then simply becomes a numerical factor corresponding to the number of SM fermions, so that $\Sigma_{\rm tot}=N_{f}=45$ \footnote{While $N_f=45$ is the correct value when all SM fermions are in equilibrium, this number is subject to change when the temperature drops sufficiently low and SM fermions with large mass decouple from the radiation bath. We numerically checked the impact of these changes in $N_f$ on the final DM abundance, and the effect is negligible in the vast majority of our parameter space, even for small reheating temperatures. We have thus omitted the evolution of $N_f$ from our analysis.}.

Following the shape of the isodensity curve in Fig.~\ref{Fig:PureVectorCoupling} for $M_{Z'}=10$ TeV
(top-left panel), we clearly distinguish five different dependencies 
$\Trh=f(m_\chi)$.
Three behaviors are within the UFO (shaded blue) framework, whereas two belong to the WIMP-like regime (shaded red).
For $m_{\chi}\approx10^{-7}$~GeV, and sufficiently large $\Trh$,
the correct relic abundance is obtained for UFO during radiation domination \cite{Henrich:2025gsd}. 
We then have $n_\chi(\Trh)\propto \Trh^3$, and from Eq.~(\ref{Eq:omegageneric}), the relic abundance is only determined by $m_\chi$.
This is the reason why each contour in the $m_{\chi}<T_{\rm RH}$ region in Fig.~\ref{Fig:PureVectorCoupling} becomes vertical near $m_{\chi}\approx 10^{-7}$~GeV \cite{Henrich:2025gsd} similar to the mass saturating the relic density bound for SM neutrinos with $\mathcal{O}(10)$~eV masses. For all the vertical part of the curve, freeze-out occurs after reheating. 
However, for each $M_{Z'}$, there exists a $\Trh$ {\it below
which} the decoupling happens {\it during} the reheating.
In this case, for higher DM masses, it is necessary to 
{\it decrease} $\Trh$ in order to increase the entropy injection period and thus dilute the dark matter more effectively, as the DM production peaks at $\Trh$.
Combining Eqs.~(\ref{Eq:yrhUFOmLeqTRH})
and (\ref{Eq:omegageneric}), we have for $m_\chi< \Trh$,
\begin{equation}
\Omega_{\chi}h^2\propto \frac{T_{\rm RH}^3 m_{\chi}}{M_{Z'}^4} \Rightarrow T_{\rm RH}|_{\Omega_{\chi}h^2=0.12}\propto\left(\frac{M_{Z'}^4}{m_{\chi}}\right)^{1/3}\text{ (for $m_{\chi}<T_{\rm RH}$)}\,,
\label{Eq:relicMleqTRH}
\end{equation}
which is the behavior reflected in the slope of each of the contours in Fig.~\ref{Fig:PureVectorCoupling} in the mass range
$10^{-7}~{\rm GeV} \lesssim m_{\chi}\lesssim T_{\rm RH}$ (the region above and to the left of the thin gray line). 
Note that the value of $\Trh\equiv \Trh^{UFO}$ required for DM to decouple during reheating,
increases with $M_{Z'}$ as $\Trh^{UFO}\propto M^\frac43_{Z'}$. This is expected from Eq.~(\ref{Eq:TFObis}) and is easily seen by comparing the lowest value of $\Trh$ at $m_\chi \sim 100$~eV at different values of $M_{Z'}$. However, for $m_\chi>\Trh$, the situation is drastically different.

Indeed, for higher masses, when $m_\chi> \Trh$,
the production of DM ends before the end of the reheating. In this case, there is more entropy produced as the reheating process continues between 
$\am$ and $\arh$. Therefore, it is necessary to raise $\Trh$ to limit the dilution of the DM abundance by increasing $(\frac{\am}{\arh})^3$. 
Combining Eqs.~(\ref{Eq:yrhUFOmGeqTRH}) and (\ref{Eq:omegageneric}), we obtain 
\begin{equation}
\Omega_{\chi}h^2\propto \frac{T_{\rm RH}^7}{M_{Z'}^4 m_{\chi}^3} \Rightarrow T_{\rm RH}|_{\Omega_{\chi}h^2=0.12}\propto\left(M_{Z'}^4 m_{\chi}^3\right)^{1/7}\text{ (for $m_{\chi}>T_{\rm RH}$)}\,,
\label{Eq:relicMGeqTRH}
\end{equation}
where we used Eq.~(\ref{arats1}) for $\am/\arh$.
This relation between $\Trh$ and $m_\chi$ is reflected in the contours of Fig.~\ref{Fig:PureVectorCoupling}
for DM masses greater than $T_{\rm RH}$ but for DM masses small enough to not undergo WIMP-like freeze-out (in the blue and purple shaded regions below and to the right of the gray line). 
For larger $m_\chi$, DM freeze-out occurs during reheating when $\chi$ is non-relativistic and the curve enters the WIMP--like (non-relativistic) freeze out regions shown by the red shaded area in Fig.~\ref{Fig:PureVectorCoupling}.

Indeed, whereas the behavior expressed in Eqs.~(\ref{Eq:relicMleqTRH}) and (\ref{Eq:relicMGeqTRH}) accurately describes the relic abundance when either FI or UFO 
is the operative production mechanism, 
these expressions no longer hold when WIMP-like freeze-out
occurs. 
For sufficiently large $m_{\chi}\gtrsim \TFO$ given by Eq.~(\ref{Eq:TFO}), 
the DM will transition from UFO to WIMP-like 
freeze-out, or from freeze-in directly to WIMP-like freeze-out.
We can determine the boundary between UFO and WIMP-like FO for $m_\chi>\Trh$ by using Eq.~(\ref{Eq:TFO}) and Eq.~(\ref{Eq:xFOboundary}), where we recall that the latter equation provides the value of $\xfo$ at the WIMP/UFO boundary. 
This implies that WIMP-like FO will occur for 
$T_{\rm RH}\gtrsim35 \text{ MeV}\left(\frac{M_{Z'}}{10^4 \text{ GeV}}\right)^{2}\left(\frac{100 \text{ GeV}}{m_\chi}\right)^{1/2}$, which is the border between 
the UFO (blue) and the WIMP-like FO (red) regions in the Fig.~\ref{Fig:PureVectorCoupling} for $m_\chi$ sufficiently larger than $T_{\rm RH}$. 
For $m_\chi < \Trh$, one should use Eq.~(\ref{Eq:TFObis}) for $\TFO$, which gives 
$m_\chi \gtrsim 2.4 ~{\rm GeV}\left(\frac{M_{Z'}}{10^4~\rm GeV}\right)^\frac43$, also
in accordance with Fig.~\ref{Fig:PureVectorCoupling}.

Moreover, as we see in the panels of Fig.~\ref{Fig:PureVectorCoupling}, there are 
two different scalings of $\Trh(m_\chi)$ within the WIMP-like region. For $m_\chi < M_{Z'}$, we can estimate the relation between $\Trh$ and $m_\chi$ taking into account the dilution between $\TFO$ and $\Trh$ so that Eq.~(\ref{Eq:omegageneric}) becomes
\beq
\Omega_{\chi}h^2\simeq5.88\times 10^6 \, \left(\frac{n_{\chi}(\TFO)\Trh^5}{\TFO^8}\right)\left(\frac{m_\chi}{1~\rm GeV}\right)\,.
\label{Eq:omegageneric2}
\eeq
Then using the non-relativistic form for  $n_\chi(\TFO)$ from Eq.~(\ref{Eq:neq}), we see that 
\beq
\Trh \sim m_\chi^{\frac45} e^{\frac{x_{\rm FO}}{5}}/x_{\rm FO}^{\frac{13}{10}} \, .
\label{wimptrh}
\eeq
While the dominant dependence of $\Trh$ on $m_\chi$ comes from the first term ($m_\chi^\frac45$), there is also some mild dependence on
both $\Trh$ and $m_\chi$ in $e^{\xfo/5}/x^{13/10}$ resulting in the nearly linear growth in $\Trh$ with respect to $m_\chi$ in the WIMP-like FO region with $m_\chi < M_{Z'}$ seen in Fig.~\ref{Fig:PureVectorCoupling}.
Note that Eq.~(\ref{Eq:omegageneric2}) is an approximate form for the relic density. All of our results in Fig.~\ref{Fig:PureVectorCoupling} are based on a numerical integration of the Boltzmann equations. 
As noted earlier for even larger values of $m_\chi\gtrsim M_{Z'}$, the production cross section changes and the solution for $x_{\rm FO}$ is given instead by Eq.~(\ref{xfo3}) in which case for example, with  $(m_\chi,\Trh,M_{Z'})=(10^6, 10^3, 10^4)$ GeV, one obtains
$x_{\rm FO} \simeq 14$. While the functional dependence of $\Trh$ given in Eq.~(\ref{wimptrh})
is still valid, the dependence on $m_\chi$ differs as the solution for $x_{\rm FO}$ has a difference dependence on $m_\chi$ for $m_\chi > M_{Z'}$, 
 This change in $x_{\rm FO}$ then results in the change of slope seen when $m_\chi > M_{Z'}$ in the upper two panels of Fig.~\ref{Fig:PureVectorCoupling}.
Note that even though $m_\chi > M_{Z'}, \Trh$,
$m_\chi < \Tmax$ and $\chi$ does enter into equilibrium with the SM bath throughout this WIMP-like freeze-out region.

We also see by comparing the panels in Fig.~\ref{Fig:PureVectorCoupling} 
that as $M_{Z'}$ is increased, the region where freeze-in occurs becomes more prominent. The regions shaded as freeze-in never come into equilibrium
with the SM.
The condition for DM entering into thermal equilibrium is  
\beq
\frac{R_\chi(M_{Z'})}{n_{\rm eq}(M_{Z'})}> \frac32 H(M_{Z'})\,.
\label{Eq:fimpfrontier}
\eeq
Indeed, it was shown in \cite{Henrich:2025gsd,HMO2} 
that if at $T=M_{Z'}$ the DM has not yet entered in  thermal 
equilibrium, it will never reach it (for matter-like reheating and a DM production process characterized by $n=2$). Using Eq.~(\ref{Eq:rategeneric}), the condition (\ref{Eq:fimpfrontier}) to ensure thermal equilibrium occurs (UFO regime) becomes, for $m_\chi < M_{Z'}$
\beq
\frac34 \frac{g_\chi \zeta(3)}{\pi^2}\beta M_{Z'} > 
\frac32 \sqrt{\frac{\alpha}{3}}\frac{M_{Z'}^4}{M_P\Trh^2} \,,
\eeq
or
\beq
\Trh \gtrsim
1.7~{\rm GeV} \left(\frac{M_{Z'}}{10^6~\rm GeV}\right)^\frac32\,.
\label{wimptrh1}
\eeq
This condition is represented by the horizontal boundaries between the UFO (blue) and FIMP (purple) regions in 
Figs.~\ref{Fig:PureVectorCoupling}.
As expected for larger $Z'$ masses, it becomes more difficult to enter into thermal equilibrium. This explains why the freeze-in region increases for larger $M_{Z'}$, and is independent of the mass of $\chi$.

For $m_\chi> M_{Z'}$, 
using $R_\chi/n^2_{\rm eq} \approx \frac{\hat{\beta}}{m_{\chi}^2}$ 
with $\hat \beta$ given by Eq.~(\ref{Eq:betahat}), the condition (\ref{Eq:fimpfrontier}) must be modified and evaluated at $\TFO$ rather than $M_{Z'}$ so that
\beq
\Trh > \left(\frac{x_{\rm FO}^{-5/2}e^{x_{\rm FO}} \sqrt{6\pi^3\alpha}}{g_\chi \hat{\beta}}\right)^\frac12 
\frac{m_\chi^{3/2}}{M_{P}^{1/2}}\,.
\label{wimptrh2}
\eeq
which for $x_{\rm FO}=13/2$ gives the approximate boundary between the WIMP and FI regions in the $m_\chi>M_{Z'}$ regime. Neither (\ref{wimptrh1}) or (\ref{wimptrh2}) are exact because decays and the resonance region also affect when the transition to equilibrium occurs, which these estimates neglect.

From Fig.~\ref{Fig:PureVectorCoupling}, we see that for $M_{Z'} = 10^4$~GeV, low mass $\chi$'s always enter into equilibrium for all $\Trh$ greater than the BBN limit. 
However for larger masses, equilibrium is not reached and there 
is a purple shaded region when $m_\chi > M_{Z'}$.  For larger $M_{Z'}$, 
equilibrium is attained for $\Trh \gtrsim 2$~GeV for $M_{Z'} = 10^6$~GeV and $\Trh \gtrsim 200$~GeV for $M_{Z'} = 10^8$~GeV (recall that the rate is proportional to $1/M_{Z'}^4$).

\begin{figure}[!ht]
\centering
\vskip .2in
\includegraphics[width=4in]{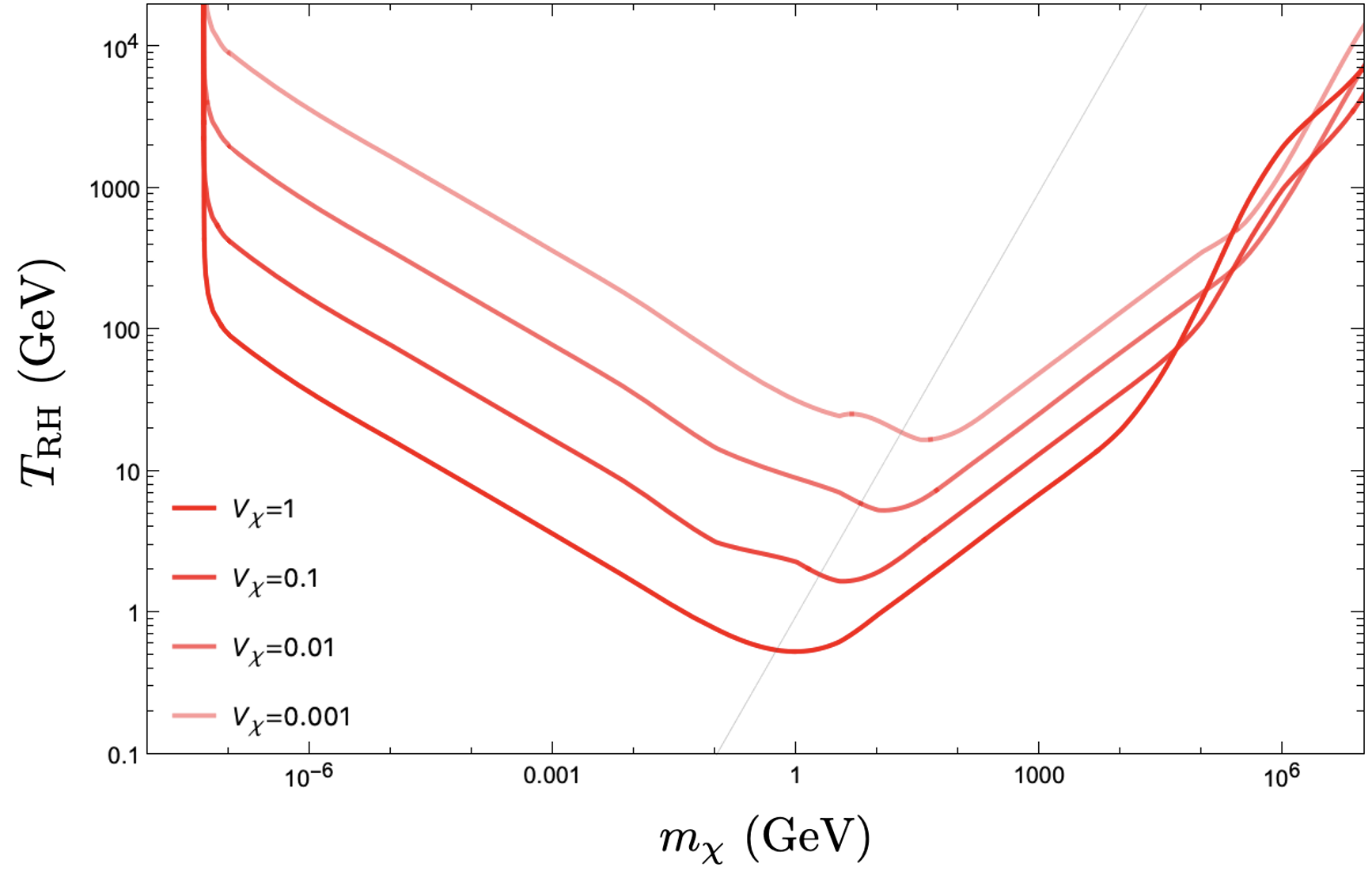}
\caption{\em \small Comparison of contours in the $(m_\chi,\Trh)$ plane for different values of the coupling $V_\chi$ for $M_{Z'}=10^6$~GeV. Each contour corresponds to the correct relic abundance $\Omega_\chi h^2=0.12$. The thin gray line corresponds to $m_\chi=\Trh$.}
\label{Fig:CouplingComparison}
\end{figure}

We show in Fig.~\ref{Fig:CouplingComparison}
the resulting $\Omega_\chi h^2 = 0.12$ contours for 
 different choices of the coupling $V_\chi$ (still keeping $A_\chi=0$) 
 and with $M_{Z'} = 10^6$~GeV.
 We see that for $m_\chi < M_{Z'}$, changing $V_\chi$ is equivalent to a rescaling\footnote{This scaling is not exact of course and is only valid in the limit $m_{f},m_{\chi} \ll T \ll M_{Z'}$ used in Eq.~(\ref{Eq:rateLimit1}) and when $Z'$ annihilations and decays can be neglected. } of $M_{Z'}$
since the dominant rates are all proportional to $(V_\chi/M_{Z'})^4$.
This scaling does not hold however when $m_\chi > M_{Z'}$. 
As one can see in Fig.~\ref{Fig:CouplingComparison}, at low $m_\chi < \Trh$, the
 scaling holds as expected. However at $m_\chi > \Trh$, the scaling is modified as the approximation used in Eq.~(\ref{Eq:rateLimit1})
 breaks down. For $m_\chi \gtrsim M_{Z'}$, this scaling is lost entirely and we see that in some cases the reheating temperature for low $V_\chi$ is required to be higher than that for $V_\chi = 1$. This change in slope in the $m_\chi>T_{\rm RH}$ region occurs at the UFO/WIMP boundary for each choice of $V_\chi$.

\subsection{Vector vs. axial vector coupling}

While Fig.~\ref{Fig:PureVectorCoupling} depicts the relic abundance for the case of a pure vector coupling ($V_{\chi}=V_{f}=1, A_{\chi}=A_{f}=0$) we can also consider a pure axial vector coupling ($A_{\chi}=A_{f}=1$, $V_{\chi}=V_{f}=0$). In this case, we obtain a contour in the $(m_{\chi},T_{\rm RH})$ plane which is quite similar to the pure vector coupling case, with some minor discrepancies  as seen in Fig.~{\ref{Fig:VectorvsAxialComparison}}.

\begin{figure}[!ht]
\centering
\vskip .2in
\includegraphics[width=6in]{images/251023_VV_AA_relic_v2.png}
\caption{\em \small Comparison of the viable parameter space for vector vs. axial vector couplings for $M_{Z'}=10^6$~GeV. Contours in the $(m_{\chi},T_{\rm RH})$ correspond to $\Omega_{\chi}h^2=0.12$. The left panel displays the entire parameter space, while the right panel is a zoom of the left plot which depicts the regime where the two coupling contours differ. The gray lines in each panel correspond to $m_{\chi}=T_{\rm RH}$.}
\label{Fig:VectorvsAxialComparison}
\end{figure}

First of all, for $m_{\chi} \ll T_{\rm RH} \ll M_{Z'}$ so that $s \gg m_{\chi},m_{f}$ we expect the vector and axial vector rates to share the same limit given in Eq.~(\ref{Eq:rateLimit1}) since 
from Eq.~(\ref{Eq:geff}), $g_{\rm eff,f} = 1$ in both cases. This is because 
the form of the squared amplitudes in Eqs.~(\ref{Eq:Mvv}) and (\ref{Eq:Maa}) 
for the two processes approach the same form. This then directly implies 
that the thermally averaged cross sections for the two processes will 
converge when $T\gg m_{\chi},m_{f}$. This is clearly seen in the left panel 
of Fig.~\ref{Fig:VectorvsAxialComparison} where the two curves are indistinguishable for $m_\chi < \Trh$. 
Recall that when $T_{\rm RH}>m_{\chi}$, the majority of the DM production 
will occur when $T\approx T_{\rm RH}$, such that production will happen when $T\gg m_{\chi},m_{f}$ for the entire parameter space above the gray line in 
Fig.~\ref{Fig:VectorvsAxialComparison}. This accounts for the blue and 
purple contours converging in the $m_{\chi}<T_{\rm RH}$ region of the left panel of Fig.~\ref{Fig:VectorvsAxialComparison}.

When $m_{\chi}>T_{\rm RH}$ and when $m_{\chi}$ is sufficiently small such that the DM does not undergo WIMP-like freeze-out (but rather UFO or freeze-in), the two contours in Fig.~{\ref{Fig:VectorvsAxialComparison}} differ 
from one another. The pure axial vector interaction is weaker than the corresponding vector interaction, especially at low temperatures. 
This can be seen by comparing Eqs.~(\ref{Eq:Mvv}) and (\ref{Eq:Maa}). Setting $m_f = 0$, we see that the amplitude squared for the axial vector 
interactions has an extra term $- 4 s m_\chi^2$, which decreases the rate and therefore requires a higher reheating temperature to obtain the same 
relic density.  Nevertheless, the two interactions are similar to one 
another and hence the total DM production is quite similar. This is 
consistent with other studies of $Z'$ portal DM which compare pure vector and pure axial couplings \cite{Arcadi:2024}. 

For still higher $m_\chi$, we transition from UFO to WIMP-like FO and
this trend is reversed so that the vector interaction requires a higher reheating temperature to obtain the same relic density. This can be understood as follows. In all cases, the vector interaction is stronger than the axial vector. For freeze-in and UFO, this corresponds to requiring a higher reheating temperature for the axial vector interaction. For WIMP-like FO, this corresponds to {\em smaller} final abundance for the vector interaction since the DM stays in equilibrium longer and more annihilations occur. Thus, the vector interaction requires higher $\Trh$ in this regime.

\begin{figure}[!ht]
\centering
\vskip .2in
\includegraphics[width=3.8in]{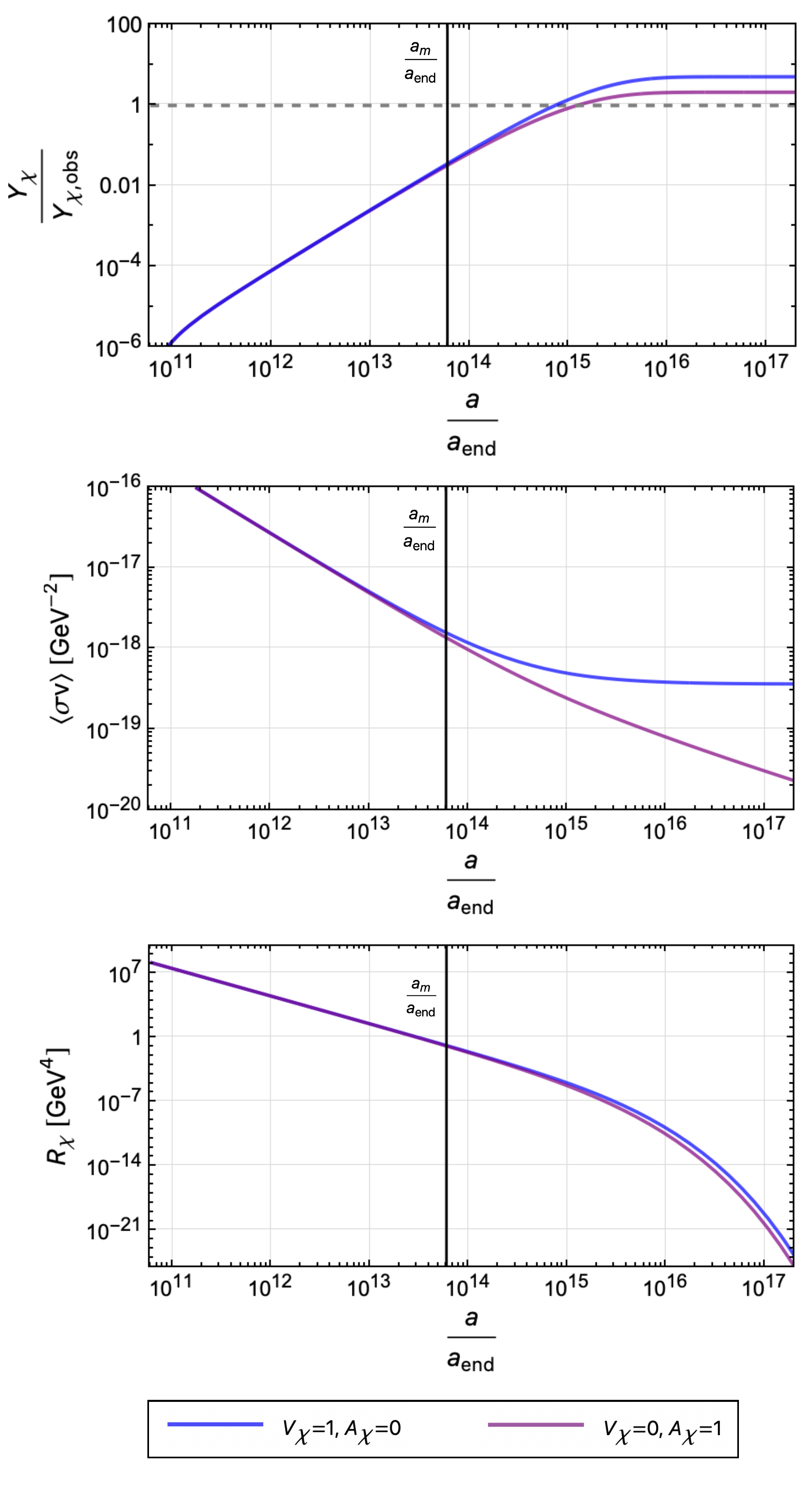}
\caption{\em \small Comparing the behavior of pure vector coupling (blue) vs. pure axial vector coupling (purple) during non-instantaneous reheating. The top panel depicts the evolution of the co-moving DM number density ($Y_{\chi}$) for each process as a function of the scale factor $a$. The middle panel depicts the respective thermally averaged cross sections, and the bottom panel depicts the respective rates $R_{\chi}$. The pure axial vector process is velocity suppressed for low $T$, relative to the pure vector coupling. The parameter choices for these plots are $M_{Z'}=10^6$~GeV, $m_{\chi}=10^{3}$~GeV, and $T_{\rm RH}=10$~GeV. The vertical black line corresponds to $\frac{a_{\rm m}}{a_{\rm end}}$, when the bath temperature is $T=m_{\chi}$.}
\label{Fig:VectorvsAxialComparisonSigmaV}
\end{figure}

The convergence of the pure vector vs.~pure axial vector processes in the high temperature regime can also be seen in Fig.~\ref{Fig:VectorvsAxialComparisonSigmaV}, where the top panel shows the evolution of the co-moving DM number density and the bottom two panels show the thermally averaged cross sections and the DM production rate, $R_{\chi}$, for the two interactions as a function of the scale factor $a$ during reheating. It is evident that when $T \gtrsim m_{\chi}$, the behavior of the vector and axial vector processes converge in all three plots. Recall that $a_{m}$ is the scale factor for which the bath temperature is $T=m_{\chi}$. Thus, the regime $T>m_{\chi}$ corresponds to $a<a_{m}$. At late times, we see the effect of the reduced squared amplitude for axial vector interactions results in a lower DM yield. 

For low
temperatures ($T \lesssim m_{\chi}$), the cross section for the pure axial vector coupling is velocity suppressed relative to the pure vector coupling, as can be clearly seen in the middle panel of Fig.~\ref{Fig:VectorvsAxialComparisonSigmaV}. This velocity suppression significantly impacts the corresponding cross sections that are relevant for DM detection efforts \cite{ Dienes:2013xya}. However, it is important to notice that while the cross sections at low temperatures are quite different from one another, the total DM production that results from the two processes is still quite similar, because the majority of the DM production happens at $T\approx m_{\chi}$, when the interaction strengths of the two processes have not yet significantly diverged. This is evident from the top panel in Fig.~\ref{Fig:VectorvsAxialComparisonSigmaV}. This lack of significant difference in the total DM production is also evident when examining the rate of DM production per unit volume ($R_{\chi}$). Because the two cross sections diverge from one another precisely in the low temperature regime when $T\lesssim m_{\chi}$, the effect of this discrepancy is blunted by the fact that both processes become Boltzmann suppressed in this regime. This can be seen by the similarity between the blue and purple curves in the bottom panel of Fig.~\ref{Fig:VectorvsAxialComparisonSigmaV}.

\subsection{Interactions with both vector and axial vector couplings}

\begin{figure}[!ht]
\centering
\vskip .2in
\includegraphics[width=4in]{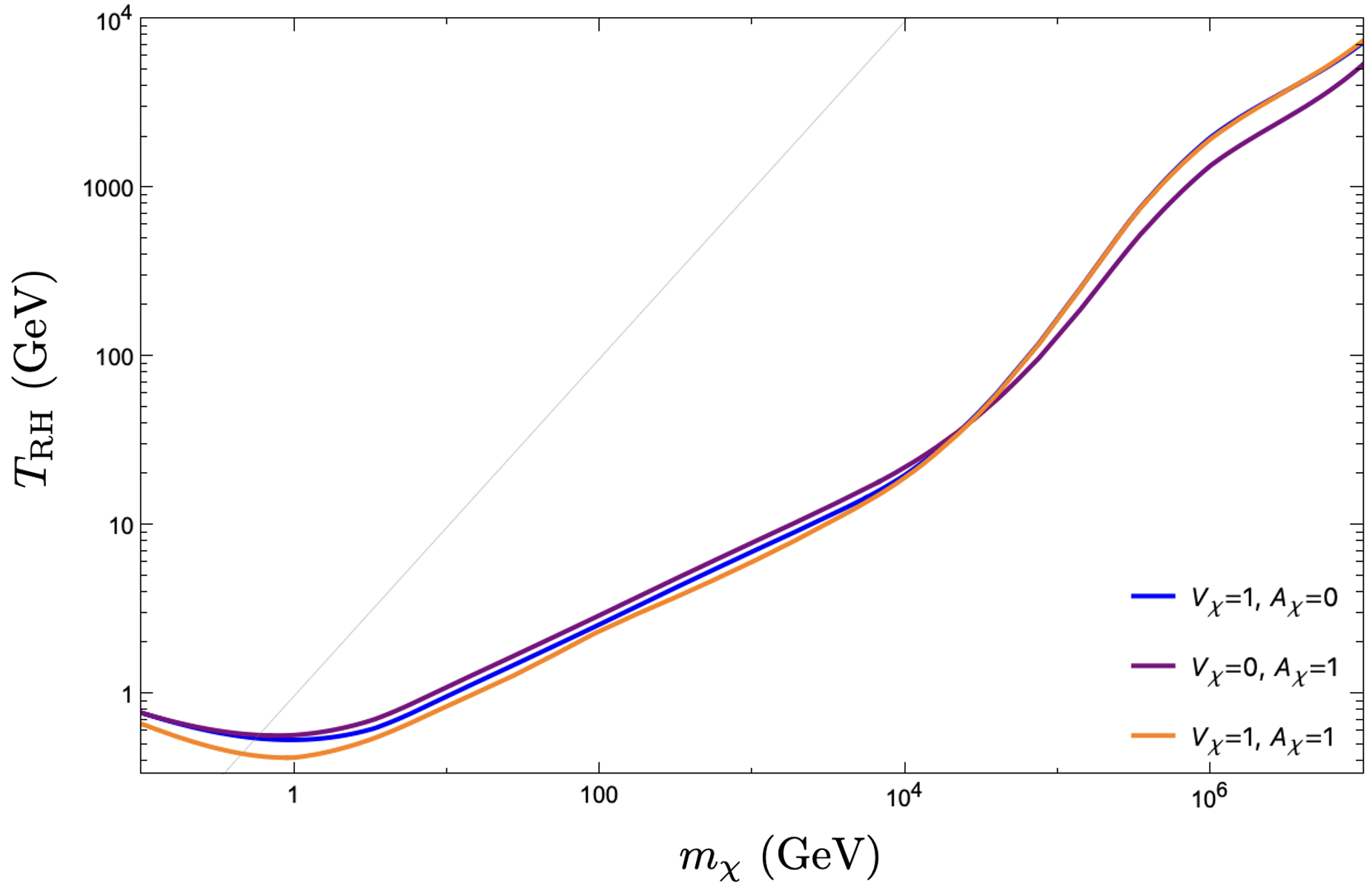}
\caption{\em \small Comparison of the viable parameter space for vector vs. axial vector couplings for $M_{Z'}=10^6$~GeV. Contours in the $(m_{\chi},T_{\rm RH})$ plane correspond to $\Omega_{\chi}h^2=0.12$. The gray line corresponds to $m_{\chi}=T_{\rm RH}$.}
\label{Fig:VV_VA_AA_comparison}
\end{figure}

In addition to the pure vector and pure axial vector cases described above, 
it is also possible to have both vector and axial vector couplings turned on, which we now consider. 
This implies a new 
contribution to the amplitude squared for this process, given by 
Eq.~(\ref{Eq:Mva}). Thus, when both vector and axial vector couplings are 
turned on, we will need $|\overline{\mathcal{M}_{\rm vv}}|^{2}$, $ |\overline{\mathcal{M}_{\rm va}}|^{2}$, and $|\overline{\mathcal{M}_{\rm aa}}|^{2}$. The resulting contour for $V_\chi=A_\chi=1$ is displayed in Fig.~\ref{Fig:VV_VA_AA_comparison} alongside the pure vector and axial 
vector cases. In the UFO and FI regimes, the effect is additive with 
respect to the pure vector and pure axial vector cases. In the WIMP regime, 
the result is essentially equivalent to the pure vector scenario. This is 
not surprising, since the contributions to $Y_{\chi}$ are additive for UFO 
and FI, but for WIMP-like FO, the strength of the pure vector 
interaction keeps the DM in equilibrium long enough for the FO abundance to 
be essentially the same regardless of whether the axial vector interaction is turned on. 

\section{Summary}
\label{summ}

While we are quite confident that dark matter exists, we are at a loss as to its identity and to its production mechanism (knowing the former may fix the latter). Light SM neutrinos with $\mathcal{O}$(10) eV masses are the original UFO candidates \cite{Gershtein:1966gg,Cowsik:1972gh,Szalay:1974jta,SS}, though these have long been excluded as DM candidates due to their detrimental effect on the perturbation spectrum for structure formation. A heavy (4th generation) neutrino with $\mathcal{O}$(GeV) mass \cite{hut} was the original CDM WIMP candidate, but these too have since been excluded from the width of the $Z$ gauge boson, and from direct detection experiments. Yet, there is no dearth of candidates for CDM WIMPs. For decades, the lightest supersymmetric particle
\cite{gold,ehnos} was a prime example of a WIMP 
whose relic density is determined by non-relativistic FO and while this remains a viable possibility \cite{Ellis:2022emx,Antusch:2025rbp}, direct detection experiments have turned up empty, as have accelerator searches for supersymmetry. Supergravity models provided a promising alternative, namely the gravitino \cite{grav,ehnos, kl}, which was the original FIMP candidate produced by freeze-in \cite{grav2}. However, these and other FIMPs are exceedingly difficult to detect due their feeble couplings. While WIMP and FIMP candidates abound, UFO candidates have been frequently neglected due to their association with hot DM.

Here, we have considered a simple BSM model consisting of a (Dirac) fermion candidate coupled to the SM through a heavy $Z'$ mediator with either (or both) vector and axial vector interactions. We have focused our analysis on the possibility that DM is produced during the non-instantaneous reheating period. Interestingly, depending on the mass of the DM candidate, $m_\chi$, and the mass of $Z'$, $M_{Z'}$, we have shown that $\chi$ may be produced via either freeze-in, (ultra)relativistic freeze-out, or non-relativistic freeze-out, so long as an appropriate reheating temperature, $\Trh$, is chosen to obtain the correct relic density. We have shown that when $M_{Z'}>T_{\rm RH},m_\chi$, UFO will be the dominant mechanism of DM production for much of the available parameter space (see e.g. Fig.~\ref{Fig:PureVectorCoupling}). For $Z'$ masses in the range $10^4 - 10^{10}$~GeV, and $m_\chi$ in the range $10^{-7}-10^7$~GeV, considered here, we require a reheating temperature between 4~MeV and $10^6$~GeV (higher reheating temperatures are suitable for $M_{Z'} \gtrsim 10^{10}$~GeV depending on $m_\chi$). The Dirac fermion DM is assumed to have relatively large gauge couplings, and as such is not a conventional FIMP, but is more like a NETDM candidate \cite{Mambrini:2013iaa,Mambrini:2015vna} with production rates potentially suppressed by the mass of the heavy $Z'$.

Our central results are given in Fig.~\ref{Fig:PureVectorCoupling} which shows the required reheating temperature as a function of the DM mass for different choices of $M_{Z'}$.  For low $M_{Z'} \lesssim 10^4$~GeV, the UFO mechanism is always operative for $m_\chi \lesssim 100$~GeV. The reheating temperature required to yield the correct DM abundance is determined by solving for the necessary dilution of $\chi$ from its freeze-out density through the end of reheating and eventually to the present day. For larger DM masses, WIMP-like FO occurs {\em during} reheating and therefore the allowed DM mass range is significantly broader than the 
conventional thermal freeze-out (occurring after reheating. 

For $M_{Z'} > 10^4$~GeV, the DM interactions with the SM can be sufficiently weak that $\chi$ never enters into equilibrium with the SM bath and the conventional FI process dominates DM production for a range of DM masses. For example, for $M_{Z'} = 10^6$~GeV, the correct relic density is obtained by UFO for $10^{-7}~{\rm GeV} < m_\chi< 10^{-3}~{\rm GeV}$ {\em and} $100~{\rm GeV} < m_\chi < 10^4~{\rm GeV}$. FI is applicable for masses between $10^{-3}~{\rm GeV}$ and $100~{\rm GeV}$, while WIMP-like FO operates for SM masses larger than $10^4~{\rm GeV}$. Thus the viable range for DM masses (with reasonably large couplings) is greatly extended by UFO.

Reheating is a necessary component of all inflation models. Often assumed to be an instantaneous process after inflation, there is a finite period in the post-inflationary evolution of the Universe in which the thermal bath is first established and lasts until the radiation comes to dominate the energy density of the Universe. Indeed, shortly after inflation ends, the thermal bath reaches a maximum temperature (typically of order $10^{12}$~GeV), and cools as radiation is continuously produced due to inflaton decays while the Universe expands.  
In this work, we concentrated on parameter choices such that $\chi$ decouples from the thermal bath during this period of reheating. In particular, we have related the three mechanisms in which 
1) the DM candidate was never in thermal equilibrium and FI production occurs, 2)
the candidate was in thermal equilibrium and $\TFO \gg m_\chi$, such that UFO production occurs or 3) the candidate was in thermal equilibrium and $\TFO \lesssim m_\chi$, and WIMP-like FO production occurs. While WIMP-like FO and FI have been studied exhaustively, UFO is an unavoidable intermediate regime between the WIMP and FIMP regimes for interactions with sufficiently steep temperature dependence \cite{HMO2}, including heavy mediator interactions. We have shown here that UFO is in fact the dominant mechanism for much of the relevant parameter space when DM production occurs during reheating, and therefore cannot be ignored. While we have here studied DM production during reheating, our general conclusions are also applicable to a non-standard cosmological era of early matter domination (EMD).

The ability to obtain the correct CDM relic density with relatively large cross sections is crucial for the detectability of DM. The fact that UFO candidates enter equilibrium and have cross sections that are significantly larger than typical FIMPs (but small enough to evade current WIMP constraints) will allow for direct detection scattering rates within the limits of existing and proposed experiments searching for DM.  This will be the focus of future work.

\acknowledgments
This project has received support from the European Union's Horizon 2020 research and innovation program under the Marie Sklodowska-Curie grant agreement No 860881-HIDDeN.
  The work of K.A.O.~was supported in part by DOE grant DE-SC0011842  at the University of
Minnesota. Y.M. acknowledges support by Institut Pascal at Université Paris-Saclay during the Paris-Saclay Astroparticle Symposium 2025, with the support of the P2IO Laboratory of Excellence (program “Investissements d’avenir” ANR-11-IDEX-0003-01 Paris-Saclay and ANR- 10-LABX-0038), the P2I axis of the Graduate School of Physics of Université Paris-Saclay, as well as the CNRS IRP UCMN.

\end{document}